\documentclass[11pt]{article}

\usepackage{jheppub} 

\usepackage{braket,slashed,bm}
\usepackage{array,multirow}
\usepackage[normalem]{ulem}
\usepackage[T1]{fontenc} 

\usepackage{mathrsfs}

\usepackage{tikz}
\usetikzlibrary{arrows,decorations.pathmorphing,backgrounds,positioning,fit,petri,automata,shadows,calendar,mindmap,
decorations.markings,calc}

\def\nn{\nonumber }

\def\abs#1{\left| #1 \right| }

\def\sc{{\overline s}}

\def\bea{\begin{eqnarray}}
\def\eea{\end{eqnarray}}

\title{
On one-loop corrections in the Standard Model Effective Field Theory; the $\Gamma(h \rightarrow \gamma \, \gamma)$ case.}

\author{
Christine Hartmann and Michael Trott,\\
Niels Bohr International Academy,
University of Copenhagen, Blegdamsvej 17, DK-2100 Copenhagen, Denmark}

\abstract{
We calculate one loop contributions to $\Gamma(h \rightarrow \gamma \, \gamma)$ from higher dimensional operators
in the Standard Model Effective Field Theory (SMEFT). Some technical challenges related to
determining Electroweak one loop "finite terms" are discussed and overcome.
Although we restrict our attention to $\Gamma(h \rightarrow \gamma \, \gamma)$, several developments we report have broad implications. Firstly, the running of the vacuum expectation value modifies the $\log(\mu)$ dependence of processes in a
manner that is not captured in some past SMEFT Renormalization Group (RG) calculations. Secondly, higher dimensional operators can source ghost interactions in $R_\xi$ gauges due to a modified gauge fixing procedure. Lastly, higher dimensional operators can contribute with pure finite terms at one loop in a manner that is not anticipated in a RG analysis. These results cast recent speculation on the nature of one loop corrections in the SMEFT in an entirely new light.}

\begin{document}
\maketitle

\section{Introduction}

The properties of the Standard Model (SM) Higgs Boson are consistent with the measured properties of the $0^+$ particle discovered in 2012. The discovery of this state was the main accomplishment of the first run of the LHC. After a two year shutdown, the beams of the LHC have started circulating once again and within a month, collisions will start providing data from Run II. The effects of beyond the Standard Model (BSM) physics will be robustly searched for in Run II,
in searches for new resonances and by studying the properties of the discovered $0^+$ state ever more precisely. It is important to systematically improve the theoretical frameworks that allow consistent interpretations of
the increasingly precise measurements of the properties of the $0^+$ state in tandem with the increasing experimental precision.  In this paper, we advance this effort by determining the complete one loop contribution to $\Gamma(h \rightarrow \gamma \, \gamma)$ due to a set of higher dimensional operators, including finite terms, in the linearly realized Standard Model Effective Field Theory (SMEFT).
The linear SMEFT assumes that the low energy limit of any BSM physics is adequately described when two requirements are fulfilled;
the observed $0^+$ scalar is embedded in the Higgs doublet, and Higher dimensional operators, denoted $O_i$ and built out of the $\rm SU(3) \times SU(2) \times U(1)$ invariant SM fields, are added to the renormalizable SM interactions.
In the remainder of this paper, all statements are to be understood to apply to the linear SMEFT.

The operator with lowest dimension to add to the SM, suppressed by a cut off scale, is of dimension five.\footnote{We neglect the effect of this operator, which we associate with neutrino mass generation and a high lepton number violating scale.} A complete classification of the dimension six operators is given in Refs.~\cite{Buchmuller:1985jz,Grzadkowski:2010es}, the latter of which finds that there are 59 (+ h.c) independent operators, assuming baryon number conservation, after eliminating redundant operators. Recently, the dimension seven operators have also been classified in Ref.~\cite{Lehman:2014jma}.

In this paper, we calculate a set of one loop contributions from dimension six operators to $\Gamma(h \rightarrow \gamma \, \gamma)$. The one loop improvement of $\Gamma(h \rightarrow \gamma \, \gamma)$ is necessary for a precise study of this process in the SMEFT. Such a calculation accounts
for the scale dependence of the operators, which is required to interpret deviations at the scale $\mu \simeq m_h$ in terms of an underlying new physics model,
matched onto at the scale $\mu \simeq \Lambda \gg m_h$.

The first step in such an analysis is to determine the Renormalization Group (RG) running of the $O_i$.\footnote{Or equivalently the Wilson coefficients of these operators. In this paper, we refer to the RG of the $O_i$ not emphasizing this minor distinction.}
For the basis in Ref.~\cite{Grzadkowski:2010es}, the required results were systematically determined in Refs.~\cite{Grojean:2013kd,Jenkins:2013zja,Jenkins:2013wua,Alonso:2013hga,Alonso:2014zka}.  Most of the $\log(\mu)$ dependence that is present in the physical impact of the $O_i$ at one loop can be directly determined in the
unbroken phase of the theory (i.e. calculating when $\langle H^\dagger \, H \rangle = 0$). We demonstrate how to determine
the remaining $\log(\mu)$ dependence.

The main focus of this paper is, however, on the calculation of the non log finite terms in $\Gamma(h \rightarrow \gamma \, \gamma)$.
The effects of the $O_i$ decouple as $\Lambda \gg m_h$.
As such, even when all the $\Lambda$ enhanced log terms are known, they are
only modestly enhanced compared to the remaining finite terms. For example, for $\Lambda \sim 3 \, {\rm TeV}$, the $\Lambda$ enhanced log terms are $\sim 6$, and larger numerical factors can occur in one loop diagrams
as pure "finite terms".  For this reason, it is essential to go beyond a minimal RG analysis when precisely interpreting future experimental bounds on $\Gamma (h \rightarrow \gamma \, \gamma)$.

Using the broken phase of the theory, (i.e. calculating when $\langle H^\dagger \, H \rangle = v^2$) to determine one loop finite terms,
is somewhat technically demanding. We show how these technical challenges are overcome
in the case of $\Gamma(h \rightarrow \gamma \, \gamma)$, and advance the full characterization of
this process at one loop in the SMEFT.  Many of our results have broad implications and inform ongoing efforts to systematically develop the SMEFT to next to leading order (NLO).\footnote{See for example the discussion in Refs.~\cite{Passarino:2012cb,Chen:2013kfa,Henning:2014wua,Englert:2014cva,Zhang:2014lya,Grober:2015cwa,sing1}.}
The summary of the most important developments are:
\begin{itemize}
\item{We explicitly demonstrate how the counterterms determined in a one loop RG analysis can be incorporated into full one loop calculations in the SMEFT, see Section~\ref{operators}.}
\item{
In some cases, we find that the one loop renormalization of the theory does not directly give the full $\log(\mu)$ dependence relevant to the one loop contribution of the $O_i$ to a physical process. When calculating assuming $\langle H^\dagger \, H \rangle = 0$,  one can identify all the operator counter terms. However,  the running of the vacuum expectation value (vev), or, equivalently, the running of $\langle H^\dagger H \rangle$, introduces further $\log(\mu)$ dependence, see Section~\ref{implications}.}
\item{We discuss how the Background Field Method (BFM) implementing $R_\xi$ gauge fixing is generalized into the case of the SMEFT, and how this leads to
the sourcing of ghosts in interactions proportional to $O_i$ Wilson coefficients, see Appendix~\ref{appendix}.}
\item{We show how a full one loop calculation can yield a "pure finite term" contribution to a process proportional to an $O_i$ Wilson coefficient, in a manner that is unanticipated in a RG analysis. This result makes clear that recent speculation on the structure of the one loop
RG in the SMEFT is not speculation on the full one loop structure of the SMEFT, see Section~\ref{implications}.}
\end{itemize}

The outline of the paper is as follows. In Section~\ref{Framework} we introduce the Effective Lagrangian, from which we calculate the finite contributions to the $\Gamma(h \rightarrow \gamma \gamma)$ decay. In this section, we also explain the renormalization scheme and we show the cancellation of the divergent contributions explicitly. In Section~\ref{finite}, the finite results are presented, including both the direct contributions from the one loop diagrams, as well as the contributions entering through the definition of the vev and through renormalization conditions defining the external fields and couplings. Subtleties regarding extra contributions through redefinitions of gauge fields and mixing angles, leading to ghost contributions, are also discussed in this section. In Section~\ref{phenomenology} some phenomenological implications are presented. We conclude in Section~\ref{conclusion}.

\section{Calculational framework}\label{Framework}
The one loop improvement of $\Gamma(h \rightarrow \gamma \, \gamma)$, due to the $O_i$ of the SMEFT, is given in part by the Effective Lagrangian
\begin{align}
\mathcal{L}^{(0)}_{6} &= C^{(0)}_{HB} \, \mathcal{O}^{(0)}_{HB} + C^{(0)}_{HW} \, \mathcal{O}^{(0)}_{HW} + C^{(0)}_{HW\!B} \, \mathcal{O}^{(0)}_{HWB} +  C^{(0)}_{W} \, \mathcal{O}^{(0)}_{W},  \nn \\
& +  C^{(0)}_{\substack{eW\\ rs}} \, \mathcal{O}^{(0)}_{\substack{eW\\ rs}} +  C^{(0)}_{\substack{eB\\ rs}} \, \mathcal{O}^{(0)}_{\substack{eB\\ rs}} + C^{(0)}_{\substack{uW\\ rs}} \, \mathcal{O}^{(0)}_{\substack{uW\\ rs}} +  C^{(0)}_{\substack{uB\\ rs}} \, \mathcal{O}^{(0)}_{\substack{uB\\ rs}}
+ C^{(0)}_{\substack{dW\\ rs}} \, \mathcal{O}^{(0)}_{\substack{dW\\ rs}} +  C^{(0)}_{\substack{dB\\ rs}} \, \mathcal{O}^{(0)}_{\substack{dB\\ rs}} + h.c.
\label{H6}
\end{align}
The operator notation used here  follows that in Ref.~\cite{Grzadkowski:2010es}, and the operators in the second line of Eqn.~\ref{H6} have Hermitian conjugates.
The indicies $r,s$ are flavour indicies. The bare operators considered in detail in this paper are normalized as
\begin{align}\label{operators2}
\mathcal{O}_{HB}^{(0)} &= g_1^2 \, H^\dagger \, H \,  B_{\mu \, \nu} \, B^{\mu \, \nu}, &
\mathcal{O}_{HW}^{(0)} &= g_2^2 \, H^\dagger \, H \,  W^a_{\mu \, \nu} \, W_a^{\mu \, \nu}, \nn \\
\mathcal{O}_{HWB}^{(0)} &= g_1 \, g_2 \, H^\dagger \, \sigma^a H \,  B_{\mu \, \nu} \, W_a^{\mu \, \nu}.
\end{align}
The SM fields and couplings are also bare on the right hand side of Eqn.~\ref{operators2}, but the $(0)$ labels are suppressed.
$\sigma^a$ are the Pauli matrices for weak isospin. The $g_i$ are the SM gauge couplings.
The gauge coupling normalization of the operators in Eqn.~\ref{operators} is chosen so that in the unbroken phase, the background field method can be more directly used to determine the counterterms.\footnote{This normalization of the operators by the gauge couplings was adopted in Ref.~\cite{Grojean:2013kd},
which is closely related to this work. The normalization chosen, however, differs by a factor of $-2$ from that used in Ref.~\cite{Grojean:2013kd}.
Conversely, the complete renormalization results of Ref.~\cite{Jenkins:2013zja,Jenkins:2013wua,Alonso:2013hga} do not introduce gauge coupling factors into the normalization of the operators.} The notation $\bar{g}_i$ is used for the canonically normalized couplings
in Ref.~\cite{Alonso:2013hga}. Although we canonically normalize the theory as in Ref.~\cite{Alonso:2013hga}, the bar notation is suppressed in this paper.
Also note that the Wilson coefficients are dimensionfull as they include a factor of $1/\Lambda^2$.

We choose to consider in detail the subset of operators
in Eqn.~\ref{operators2}, as these operators illustrate the basic issues involved with determining the Electroweak finite terms in the SMEFT.
In particular, the operator $O_{HWB}$, which leads to a redefinition of the mixing angles and $Z,A$ fields in the SM, generates most of the
technical challenges introduced in the SMEFT.\footnote{The Wilson coefficient of this operator corresponds to the $S$ parameter,
and the effects that we will discuss are present in any operator basis. In some alternative operator basis, the $S$ parameter is related to a sum of operators, which enhances the challenges involved in developing the SMEFT to NLO.}
Of course, a full one loop improvement, including all $O_i$ allowed at one loop, is eventually required. Further, the effect of redefining the input parameters of the SM prediction
of $\Gamma(h \rightarrow \gamma \, \gamma)$ is required for a complete one loop treatment of this process in the SMEFT. Our results are a first step in this direction, when
considering finite terms.

\subsection{Operator counterterms}\label{operators}

The operators are renormalized through introducing a renormalization matrix $Z_{ij}$ to cancel the extra operator induced divergences, such that
\bea
\mathcal{O}_i^{(0)} = Z_{i,j} \, \mathcal{O}_j^{(r)},
\eea
and we restrict our attention to $i,j = HB,HW,HWB$ in this paper. Here, the superscript $(r)$ is used to denote a renormalized operator.
With the normalization in Eqn.~\ref{operators2}, the one loop operator counterterms are
\bea
Z_{i,j}
 = \delta_{i,j}  + \frac{\mathcal{Z}_{i,j}}{16 \, \pi^2 \, \epsilon} \, ,
 \eea
where the matrix $\mathcal{Z}_{i,j}$ (for $d = 4 - 2 \, \epsilon$ dimensions in the $\rm \overline{MS}$ scheme) is~\cite{Grojean:2013kd}
\bea\label{countertermmatrix}
\mathcal{Z}_{i,j} = \left( \begin{array}{ccc}
 \frac{g_1^2}{4} - \frac{9 g_2^2}{4} + 6 \lambda + Y & 0 &  g_1^2 \\
  0  & -\frac{3 g_1^2}{4} - \frac{5 g_2^2}{4} + 6 \lambda + Y & g_2^2  \\
  \frac{3 g_2^2}{2}  &  \frac{g_1^2}{2}  & -\frac{g_1^2}{4} + \frac{9 g_2^2}{4} + 2 \lambda + Y \\
 \end{array}
\right),
\eea
in the basis of the operators given by $\mathcal{O}_i = \left(\mathcal{O}_{HB}, \, \mathcal{O}_{HW}, \, \mathcal{O}_{HWB} \right)$. Our notation is such that
\begin{align}
Y &= \text{Tr}\left[N_c Y_u^\dagger Y_u + N_c Y_d^\dagger Y_d + Y_e^\dagger Y_e\right] \approx N_c y_t^2,
\end{align}
while the Higgs potential is defined so that $v \sim 246$~GeV and $m_h^2=2 \, \lambda \, v^2$.
The renormalized interactions are introduced as
\bea
\mathcal{L}_6^{(0)} &=& Z_{SM} \, Z_{i,j} \, C_{i} \, \mathcal{O}^{(r)}_{j}, \nn \\
&=&  Z_{SM}  \, \Big( \mathcal{N}_{HB} \, \mathcal{O}_{HB}^{(r)} +      \mathcal{N}_{HW} \, \mathcal{O}_{HW}^{(r)} +   \mathcal{N}_{HWB}  \, \mathcal{O}_{HWB}^{(r)} \Big) .
\eea
The operators are multiplied by a factor $\mathcal{N}_i$ that includes the corrections from the counterterm matrix $\mathcal{Z}$ and absorbs coupling factors.
The $\mathcal{N}$'s are given by
\bea
\mathcal{N}_{HB} &=& \frac{1}{16 \, \pi^2 \, \epsilon} \left[\left(16 \, \pi^2 \, \epsilon + \frac{g_1^2}{4} - \frac{9 g_2^2}{4} + 6 \lambda + Y \right)  \, C_{HB}(\Lambda) + \frac{3 \, g_2^2}{2}  \, C_{HWB}(\Lambda) \right], \\
\mathcal{N}_{HW} &=&  \frac{1}{16 \, \pi^2 \, \epsilon} \left[\left(16 \, \pi^2 \, \epsilon - \frac{3 g_1^2}{4} - \frac{5 g_2^2}{4} + 6 \lambda + Y \right) \, C_{HW}(\Lambda) + \frac{g_1^2}{2} \,  C_{HWB}(\Lambda) \right] , \nn \\
\mathcal{N}_{HWB} &=&  \frac{1}{16 \, \pi^2 \, \epsilon} \left[\left(16 \, \pi^2 \, \epsilon - \frac{g_1^2}{4} + \frac{9 g_2^2}{4} + 2 \lambda + Y \right) \, C_{HWB}(\Lambda) + g_1^2 \, C_{HB}(\Lambda)
+ g_2^2 \,  C_{HW}(\Lambda)\right]. \nn
\eea
Although the subtraction required due to the introduction of the $\mathcal{O}_i$ has been performed to render the theory finite,
the subtractions to renormalize the SM interactions are still required. This is indicated by the inclusion of $Z_{SM}$ in the above expressions. The exact forms of the required $Z_{SM}$ and $Z_{i,j}$ depend on the normalization and scheme chosen.\footnote{Once the counterterm matrix is determined in the unbroken phase of the theory, it can be directly used in the broken phase.
This can be done even if the background field method is not used to render the SMEFT finite, as the one loop operator counterterms so obtained are gauge independent.
This is fortunate, as directly determining the counterterm matrix in the broken phase of the theory is difficult.}

\subsection{Background field method} \label{BFmethod}

The renormalization scheme we use is to define the one loop finite terms in  $\Gamma(h \rightarrow \gamma \, \gamma)$ in a manner that is consistent with implementing the $\rm \overline{MS}$ scheme, while simultaneously utilizing the Background Field method (BFM). In the BFM~\cite{'tHooft:1975vy,DeWitt:1967ub,Abbott:1981ke}, fields are split into classical and quantum components,
and  a gauge fixing term is added that maintains the gauge invariance of the classical background fields, while breaking the gauge
invariance of the quantum fields. We use $R_\xi$ gauge with background gauge fixing, with $\phi^\pm, \phi_0$, the Goldstone bosons defined through the convention
\bea \label{Hexpansion}
 H = \frac{1}{\sqrt{2}}\left( \begin{array}{c}
\sqrt{2} i \phi^+ \\
 h + v + \delta v + i\phi_0 \end{array}  \right).
 \eea
The one loop vev ($v+\delta v$) is defined as the classical background field expectation value for which the one point $h$ function vanishes. This induces finite (and gauge dependent) terms into the definition of the vev, which are discussed further in Section~\ref{finite}. The requirement to include these terms is indicated by the introduction of $\delta v$ in the above expression, which is formally of one loop order.

 Technical simplifications result from the use of the BFM. One can choose a gauge in the quantum calculation to one's advantage. Also, the counterterms accounting for the renormalization of the SM gauge fields and couplings cancel for a specific choice of operator normalization. This follows from the unbroken Ward identities of the theory, when using the BFM. The Ward identities result in the following relations among the SM counterterms~\cite{Denner:1994xt},67
\begin{align} \label{BFgoodness}
Z_A Z_e &= 1, &
Z_h &= Z_{\phi_\pm} = Z_{\phi_0}, &
Z_W Z_{g_2} &= 1.
\end{align}
The gauge fixing is undertaken in the BFM in a manner discussed in Ref.~\cite{Einhorn:1988tc,Denner:1994xt}.\footnote{Some implications of the BFM and standard t'Hooft gauge fixing differ in the SM compared to the SMEFT. We discuss some of these subtleties in the Appendix.}
The renormalized fields (where $F = {A,W,Z}$ and $S = h, \phi^0,\phi^\pm$) and couplings (denoted $c = e, g_2, g_1$) are defined in terms of bare fields and couplings via
\begin{align}
F_\mu &= \frac{1}{\sqrt{Z_{F}}} \, F^{(0)}_\mu, & S &= \frac{1}{\sqrt{Z_s}} \, S^{(0)},  & c &= \frac{1}{Z_c} \mu^{- \epsilon/2} c^{(0)}.
\end{align}
The factor $\mu^{- \epsilon/2}$ is included in the coupling relation to render the renormalized coupling dimensionless~\cite{Manohar:2000dt}.
The vev is also renormalized with the inclusion of $v = v^{(0)}/\sqrt{Z_v}$.
All the subtractions in the SM are defined in the $\rm \overline{MS}$ scheme for $d = 4 - 2 \, \epsilon$ dimensions.

\begin{figure}[!h]
\hspace{0.2cm}
\begin{tikzpicture}

\draw [thick] [dashed] (0,0) circle (0.75);

\filldraw (0.65,-0.1) rectangle (0.85,0.1);

\filldraw (180:0.75) circle (0.075);

\draw[decorate,decoration=snake] (0:0.75) -- +(30:1.30) ;
\draw[decorate,decoration=snake] (0:0.75) -- +(330:1.30) ;


\draw  [dashed] [thick] (-1.7,0) -- (-0.75,0);

\draw (0,-1.7) node [align=center] {(a)};

\node [left][ultra thick] at (-1.7,0) {$h$};
\node [above][ultra thick] at (0,0.8) {$h, \phi_0$};
\node [below][ultra thick] at (0,-0.8) {$h, \phi_0$};
\node [right][ultra thick] at (1.80,0.7) {$g$};
\node [right][ultra thick] at (1.80,-0.7) {$g$};

\end{tikzpicture}
\hspace{0.6cm}
\begin{tikzpicture}

\draw [thick] [dashed] (0,0) circle (0.75);

\filldraw (0.65,-0.1) rectangle (0.85,0.1);

\filldraw (180:0.75) circle (0.075);

\draw[decorate,decoration=snake] (0:0.75) -- +(30:1.30) ;
\draw[decorate,decoration=snake] (0:0.75) -- +(330:1.30) ;
\draw  [->][ultra thick]  (90:0.75)  -- + (0:0.01) ;
\draw  [->][ultra thick]  (270:0.75)  -- + (0:-0.01) ;


\draw  [dashed] [thick] (-1.7,0) -- (-0.75,0);

\draw (0,-1.7) node [align=center] {(b)};

\node [left][ultra thick] at (-1.7,0) {$h$};
\node [above][ultra thick] at (0,0.8) {$\phi^{\pm}$};
\node [below][ultra thick] at (0,-0.8) {$\phi^{\pm}$};
\node [right][ultra thick] at (1.80,0.7) {$g$};
\node [right][ultra thick] at (1.80,-0.7) {$g$};

\end{tikzpicture}
\hspace{0.6cm}
\begin{tikzpicture}

\filldraw (-0.1,-0.1) rectangle (0.1,0.1);

\draw  [dashed] [thick] (-1.2,0) -- (0.0,0);


\draw[decorate,decoration=snake] (0:0) -- +(30:1.40) ;
\draw[decorate,decoration=snake] (0:0) -- +(330:1.40) ;

\node [left][ultra thick] at (-1.3,0) {$h$};
\node [right][ultra thick] at (1.2,0.95) {$g$};
\node [right][ultra thick] at (1.2,-0.95) {$g$};

\draw (0,-1.7) node [align=center] {(c)};

\end{tikzpicture}

\caption{Diagrams contributing to $H \rightarrow g g$ decay from $\mathcal{O}_{GG}$.}
\label{diagrams}

\end{figure}
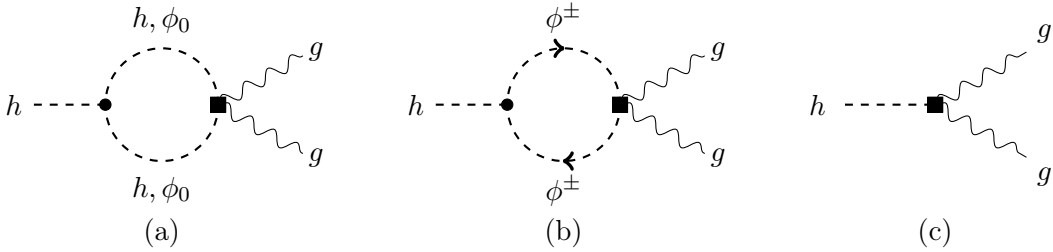

 The remaining SM counterterms required to render the theory finite are well known. The Higgs wavefunction renormalization in the background field method with the gauge fixing in Eqn.~\ref{GFresult} is
given by\footnote{The gauge coupling normalization convention used here is the same as in Ref.~\cite{Grojean:2013kd,Jenkins:2013zja,Jenkins:2013wua,Alonso:2013hga}. Another convention used in the literature
in Refs.~\cite{Machacek:1983tz,Arason:1991ic} has an alternate Hypercharge normalization convention.}
\begin{align}\label{remainingSMsubtract}
Z_h &= 1  + \frac{(3 + \xi)\, (g_1^2+ 3 \, g_2^2)}{64 \, \pi^2 \, \epsilon}  - \frac{Y}{16 \, \pi^2 \, \epsilon}.
\end{align}
The sum of the divergent terms in the vev renormalization constant $Z_v$ and $\delta v$ also needs to be fixed. A simple approach to determine this divergence that avoids operator mixing complications is to consider the operator $H^\dagger \, H G^{\mu \, \nu} G_{\mu \, \nu}$. The counterterm associated with this operator is~\cite{Grojean:2013kd}
\bea
Z_{HG} = 1 + \frac{1}{16 \, \pi^2 \, \epsilon} \left[- \frac{3 g_1^2}{4} - \frac{9 g_2^2}{4}  + 6  \lambda + Y \right].
\eea
Considering the one loop $h \rightarrow g g$ decay with the diagrams in Fig.~\ref{diagrams}, this fixes $(\sqrt{Z_v} + \frac{\delta v}{v})_{div}$ to be
\bea\label{ambiguity}
(\sqrt{Z_v} + \frac{\delta v}{v})_{div} = 1  + \frac{(3 + \xi)\, (g_1^2+ 3 \, g_2^2)}{128 \, \pi^2 \, \epsilon}  - \frac{Y}{32 \, \pi^2 \, \epsilon}.
\eea
The left hand side of Eqn.~\ref{ambiguity} reflects an ambiguity that remains in Electroweak perturbation theory when using the $\rm \overline{MS}$ scheme at one loop in this way. This requires the renormalization condition to fix $\delta v$, as discussed above. We choose to adopt the vev defining condition described in Ref.~\cite{Denner:1994xt}, which fixes the finite terms of the vev as in Section~\ref{finite}.

The renormalizations of the vev and the Higgs field are seen to satisfy the following relation,
\bea
 (\sqrt{Z_v} + \frac{\delta v}{v})_{div} = \sqrt{Z_h}.
\eea
This is in fact expected in the BFM, see Ref.~\cite{Sperling:2013eva}.

\subsection{The Effective Lagrangian}

Expanding the $O_i$ into the mass eigenstate fields, one finds the renormalized effective interactions relevant to the one loop improvement of
$\Gamma(h \rightarrow \gamma \, \gamma)$.\footnote{The distinction between the net Effective Lagrangian of the SMEFT in terms of mass eigenstate fields (Eqn.~\ref{interactions}), and an operator basis of {\it linearly independent} dimension six operators has been
unfortunately de-emphasized in some recent literature. Obscuring this important distinction makes calculations in the SMEFT beyond tree level systematically more challenging. This also
leads to the concept of "dependent" and "independent" couplings, when discussing the neologism, the "Higgs Basis" in Refs.~\cite{HiggsNonBasis,Falkowski:2015fla}.}
In all of the $\mathcal{L}_{eff}$ terms, other than some forms given in the first line of Eqn.~\ref{interactions}, the contribution to $\Gamma(h \rightarrow \gamma \, \gamma)$ decay is itself a one loop contribution.  As such, the relevant $\mathcal{L}_{eff}$ for the one loop $\Gamma(h \rightarrow \gamma \, \gamma)$ result simplifies to $\mathcal{O}(v)$ in the following way
%
\bea\label{interactions}
\mathcal{L}_{eff}(v^0,v) &=& \frac{1}{2}(h^2  + 2 \, h \, \sqrt{Z_h} \, (v \, \sqrt{Z_v} + \delta v) + \phi_0^2) \, \left(\mathcal{N}_{HB}  + \mathcal{N}_{HW} - \mathcal{N}_{HWB} \right) e^2 \, A_{\mu\nu}A^{\mu\nu}, \nn \\
&+&(\phi_+\phi_-)  \,  \left(C_{HB}  +  C_{HW}  +  C_{HWB} \right)  e^2 \, A_{\mu\nu}A^{\mu\nu}
+  (2 \, h \, v) \,  C_{HW} \, g_2^2 \, W^+_{\mu\nu} \, W_-^{\mu\nu}, \nn \\
&-&  2 \, i \, e \, g_2^2 \, ( 2 \, h \, v) \, C_{HW} \, \left[ A_{\mu} W^-_{\nu} \, W_+^{\mu\nu} + A_{\nu} W^+_{\mu} \, W_-^{\mu\nu} - A_{\mu \, \nu} W^+_{\mu} \, W_-^{\nu} \right], \nn \\
&+& 2 \, e^2 \, g_2^2 \, ( 2 \, h \, v) \, C_{HW} \,  \left(  W^+_{\mu} \, W_-^{\mu}\, A_{\nu} \, A^{\nu} -  W^+_{\mu} \, W_-^{\nu}\, A_{\nu} \, A^{\mu}\right), \nn \\
&-&  i \, e \, C_{HWB} \, g_2 \, \left[ g_2 \, ( 2 \, h \, v) \,  \left( A_{\mu \, \nu} W^+_{\mu} \, W_-^{\nu}\right)
-  (v + h) \,  A^{\mu \, \nu} (\phi_+ \, W^-_{\mu \, \nu}  - \phi_- \, W^+_{\mu \, \nu}) \right], \nn \\
&+& C_{HWB}\, v \, e^2 \, g_2 \, \Bigg(A_{\mu}(\phi^+ W_{\nu}^- + \phi^- W_{\nu}^+) - (\phi^+W_{\mu}^- + \phi^- W_{\mu}^+)A_{\nu}\Bigg)\,A^{\mu\nu}.
\eea

Each $\mathcal{N}_i$ in $\mathcal{L}_{eff}$ now contains dependence on the counterterms $Z_{g_1}$, $Z_{g_2}$, etc. implicitly. In the same way, the fields and couplings of lines two through six in $\mathcal{L}_{eff}$ should be multiplied by their respective renormalization constants. However, these corrections are two loop order and neglected. This expression is also simplified using the relations given in
Eqn.~\ref{BFgoodness}.
All of the operator Wilson coefficients in this expression are evaluated at the scale $\mu = \Lambda$, which allows a direct interpretation of any measured deviation
in terms of the underlying BSM physics sector. The combination of Wilson coefficients given by
\bea
C_{\gamma \, \gamma} = C_{HB} + C_{HW} - C_{HWB},
\eea
corresponds to the effective Wilson coefficient for $h \rightarrow \gamma \, \gamma$ at tree level and one scale. This effective Wilson coefficient, however, does not physically correspond to this process in a scale independent fashion in the SMEFT, see Section~\ref{implications}.

In addition to the terms in Eqn.~\ref{interactions}, contributions proportional to $v^2$ are present in $\mathcal{L}_{eff}$. These terms are not expected to contribute with divergent terms to the amplitude before renormalization, as the $O_i$ counterterm subtraction is proportional to $v$.\footnote{This is true at one loop.} We find
by explicit calculation that this is true.

However, finite terms could exist at one loop in $\Gamma(h \rightarrow \gamma \, \gamma)$ due to $v^2$ terms in $\mathcal{L}_{eff}$.
We find by explicit calculation that such terms do contribute at one loop. Terms of order $v^2$ come about due to the direct expansion of the operators present and from the canonical normalization of the SMEFT. Interestingly, terms of this form involve ghost fields (denoted $u^\pm$), which come about due to the effects of canonically normalizing the SMEFT. The relevant $\mathcal{O}(v^2)$ terms in $\mathcal{L}_{eff}$ are
\bea\label{interactions2}
\mathcal{L}_{eff}(v^2) &=&   2 \,  e^4  \, v^2  \, C_{HWB} \, \left( A_{\mu}A^{\mu} W_{\nu}^+W_-^{\nu} - A_{\mu}A_{\nu}W_+^{\mu}W_-^{\nu} \right) -  i \, e \, g_2^2 \, v^2 \, C_{HWB} \, A_{\mu \, \nu} W^+_{\mu} \, W_-^{\nu}, \nn \\
&+&   i\, \frac{e^3 \, v^2 \,  C_{HWB}}{\xi}\, \left( A_{\mu}W_+^{\mu}\partial_{\nu}W_-^{\nu} - A_{\mu} W_-^{\mu} \partial_{\nu}W_+^{\nu} + 2 \, e \, W^+_{\mu} \, W_-^{\nu}\, A_{\nu} \, A^{\mu} \right),
\nn \\
&+&
i \, e^3 \, v^2 \, C_{HWB}  \left( A_{\mu}W_{\nu}^+ W_-^{\mu\nu} - A_{\mu}W_{\nu}^- W_+^{\mu\nu} + W_{\mu}^+ \, W_{\nu}^- \, A^{\mu\nu}\right), \nn \\
&+& e^3 \, v^2 \,  C_{HWB}  \,  \left( i \, \phi_- \, A_{\mu} \, \partial^{\mu}\phi_+ - i \, \phi_+ \, A_{\mu} \, \partial^{\mu} \phi_-  - 2  \, e  \, \phi_+ \, \phi_- \, A_{\mu} \, A^{\mu} \right), \nn \\
& - & \frac{1}{2}  \, e^3 \, g_2 \, v^2 \,  C_{HWB}  \, A_{\mu} \, \left( W_-^{\mu} \phi_+\phi_0 + W_+^{\mu} \phi_- \phi_0 \right) -  2 \, e^4  \, v^2 \, C_{HWB} \,  A_{\mu}\, A^{\mu} \left( \bar{u}^+ \, u^+ + \bar{u}^- \, u^- \right), \nn \\
&+& i e^3  v^2   C_{HWB} \left(   \left(  \bar{u}^+ A^{\mu} \,\partial_{\mu} u^+  -  \bar{u}^- A^{\mu} \,\partial_{\mu} u^-\right)
 +     \, \partial_{\mu} A^{\mu}  \left( \bar{u}^+  u^+ - \bar{u}^-  u^- \right) \right).
\eea

\subsection{Cancelation of the divergent terms}\label{divterms}
\begin{figure}[!h]
\begin{tikzpicture}

\draw [thick] [dashed] (0,0) circle (0.75);

\filldraw (0.65,-0.1) rectangle (0.85,0.1);

\filldraw (180:0.75) circle (0.075);

\draw[decorate,decoration=snake] (0:0.75) -- +(30:1.30) ;
\draw[decorate,decoration=snake] (0:0.75) -- +(330:1.30) ;


\draw  [dashed] [thick] (-1.7,0) -- (-0.75,0);

\draw (0,-1.7) node [align=center] {(a)};

\node [left][ultra thick] at (-1.7,0) {$h$};
\node [above][ultra thick] at (0,0.8) {$h, \phi_0$};
\node [below][ultra thick] at (0,-0.8) {$h, \phi_0$};
\node [right][ultra thick] at (1.80,0.7) {$\gamma$};
\node [right][ultra thick] at (1.80,-0.7) {$\gamma$};

\end{tikzpicture}
\begin{tikzpicture}

\draw [thick] [dashed] (0,0) circle (0.75);

\filldraw (0.65,-0.1) rectangle (0.85,0.1);

\filldraw (180:0.75) circle (0.075);

\draw[decorate,decoration=snake] (0:0.75) -- +(30:1.30) ;
\draw[decorate,decoration=snake] (0:0.75) -- +(330:1.30) ;
\draw  [->][ultra thick]  (90:0.75)  -- + (0:0.01) ;
\draw  [->][ultra thick]  (270:0.75)  -- + (0:-0.01) ;


\draw  [dashed] [thick] (-1.7,0) -- (-0.75,0);

\draw (0,-1.7) node [align=center] {(b)};

\node [left][ultra thick] at (-1.7,0) {$h$};
\node [above][ultra thick] at (0,0.8) {$\phi^{\pm}$};
\node [below][ultra thick] at (0,-0.8) {$\phi^{\pm}$};
\node [right][ultra thick] at (1.80,0.7) {$\gamma$};
\node [right][ultra thick] at (1.80,-0.7) {$\gamma$};

\end{tikzpicture}
\begin{tikzpicture}

\draw  [decorate,decoration=snake] (0,0) circle (0.75);

\filldraw (-0.70,-0.1) rectangle (-0.90,0.1);

\filldraw (180:0.75) circle (0.075);

\draw[decorate,decoration=snake] (0:0.75) -- +(30:1.30) ;
\draw[decorate,decoration=snake] (0:0.75) -- +(330:1.30) ;
\draw  [->][ultra thick]  (90:0.85)  -- + (0:0.01) ;
\draw  [->][ultra thick]  (270:0.65)  -- + (0:-0.01) ;


\draw  [dashed] [thick] (-1.7,0) -- (-0.75,0);

\draw (0,-1.7) node [align=center] {(c)};

\node [left][ultra thick] at (-1.7,0) {$h$};
\node [above][ultra thick] at (0,0.8) {$W^{\pm}$};
\node [below][ultra thick] at (0,-0.8) {$W^{\pm}$};
\node [right][ultra thick] at (1.80,0.7) {$\gamma$};
\node [right][ultra thick] at (1.80,-0.7) {$\gamma$};

\end{tikzpicture}
\begin{tikzpicture}

\draw  [dashed] [thick] (-1.7,0) -- (-0.75,0);

\draw [decorate,decoration=snake] (0.5,0) arc [radius=0.9, start angle=50, end angle= 130];

\draw [decorate,decoration=snake] (-0.75,0) arc [radius=0.9, start angle=230, end angle= 310];

\draw[decorate,decoration=snake] (180:0.75) -- +(50:1.6) ;
\draw[decorate,decoration=snake] (180:0.75) -- +(310:1.6) ;


\filldraw (-0.65,-0.1) rectangle (-0.85,0.1);

\draw  [->][ultra thick]  (90:0.25)  -- + (0:0.01) ;
\draw  [->][ultra thick]  (270:0.25)  -- + (180:0.1) ;

\draw (0,-1.7) node [align=center] {(d)};

\node [left][ultra thick] at (-1.7,0) {$h$};
\node [right][ultra thick] at (0.5,0) {$W^{\pm}$};
\node [right][ultra thick] at (0.3,1.2) {$\gamma$};
\node [right][ultra thick] at (0.3,-1.4) {$\gamma$};

\end{tikzpicture}
\begin{tikzpicture}

\hspace{0.3cm}

\draw [decorate,decoration=snake] (0.7,0) arc [radius=0.7, start angle=0, end angle= 180];

\draw [dashed] (-0.75,0) arc [radius=0.7, start angle=180, end angle= 360];

\filldraw (0.6,-0.1) rectangle (0.8,0.1);
\filldraw (180:0.75) circle (0.075);

\draw[decorate,decoration=snake] (0:0.75) -- +(30:1.30) ;
\draw[decorate,decoration=snake] (0:0.75) -- +(330:1.30) ;
\draw  [->][ultra thick]  (90:0.8)  -- + (0:-0.1) ;
\draw  [->][ultra thick]  (270:0.7)  -- + (0:0.01) ;


\draw  [dashed] [thick] (-1.7,0) -- (-0.75,0);

\draw (0,-1.7) node [align=center] {(e)};

\node [left][ultra thick] at (-1.7,0) {$h$};
\node [above][ultra thick] at (0,0.8) {$W^{\pm}$};
\node [below][ultra thick] at (0,-0.8) {$\phi^{\pm}$};
\node [right][ultra thick] at (1.80,0.7) {$\gamma$};
\node [right][ultra thick] at (1.80,-0.7) {$\gamma$};

\end{tikzpicture}
\begin{tikzpicture}

\hspace{-0.3cm}

\draw  [dashed] [thick] (-1.7,0) -- (-1.2,-0.05);
\draw  [thick][dashed] (-1.2,-0.05) -- (-0.5,-0.10);

\draw [decorate,decoration=snake] (0.5,0.8) arc [radius=0.8, start angle=60, end angle= 195];

\draw [decorate,decoration=snake] (-0.5,-0.2) arc [radius=0.8, start angle=250, end angle= 380];

\draw[decorate,decoration=snake] (65:0.9) -- +(0:1.1) ;
\draw[decorate,decoration=snake] (200:0.6) -- +(335:1.7) ;

\draw  [->][ultra thick]  (110:1)  -- + (30:0.01) ;
\draw  [->][ultra thick]  (330:0.3)  -- + (210:0.01) ;

\filldraw (-0.6,-0.2) rectangle (-0.8,0);

\draw (0,-1.7) node [align=center] {(f)};

\node [left][ultra thick] at (-1.7,0) {$h$};
\node [above][ultra thick] at (-0.7,0.8) {$W^{\pm}$};
\node [below][ultra thick] at (1,0.2) {$W^{\pm}$};
\node [right][ultra thick] at (1.5,0.8) {$\gamma$};
\node [right][ultra thick] at (1,-1) {$\gamma$};

\end{tikzpicture}
\begin{tikzpicture}

\draw  [dashed] [thick] (-1.7,0) -- (-0.75,0);

\draw[dashed] (180:0.75) -- +(30:1.75) ;
\draw  [->][ultra thick]  (70:0.55)  -- + (30:0.01) ;

 \draw[decorate,decoration=snake] (180:0.75) -- +(330:1.75) ;
\draw  [->][ultra thick]  (290:0.55)  -- + (330:-0.01) ;

\draw[dashed] (50:1.2) -- +(270:1.75) ;
\draw  [->][ultra thick]  (0:0.8)  -- + (270:0.01) ;

\draw[decorate,decoration=snake] (50:1.2) -- +(0:1.1) ;
\draw[decorate,decoration=snake] (310:1.2) -- +(0:1.1) ;

\filldraw (0.65,-0.9) rectangle (0.85,-0.7);
\filldraw (180:0.75) circle (0.075);

\draw (0,-1.7) node [align=center] {(g)};

\node [left][ultra thick] at (-1.7,0) {$h$};
\node [above][ultra thick] at (0,0.8) {$\phi^{\pm}$};
\node [below][ultra thick] at (0,-0.8) {$W^{\pm}$};
\node [right][ultra thick] at (1,0) {$\phi^{\pm}$};
\node [right][ultra thick] at (1.8,0.9) {$\gamma$};
\node [right][ultra thick] at (1.8,-0.95) {$\gamma$};

\end{tikzpicture}
\begin{tikzpicture}

\draw  [dashed] [thick] (-1.7,0) -- (-0.75,0);

\draw[dashed] (180:0.75) -- +(30:1.75) ;
\draw  [->][ultra thick]  (70:0.55)  -- + (30:0.01) ;

 \draw[decorate,decoration=snake] (180:0.75) -- +(330:1.75) ;
\draw  [->][ultra thick]  (290:0.55)  -- + (330:-0.01) ;

\draw[decorate,decoration=snake] (50:1.2) -- +(270:1.75) ;
\draw  [->][ultra thick]  (0:0.8)  -- + (270:0.01) ;

\draw[decorate,decoration=snake] (50:1.2) -- +(0:1.1) ;
\draw[decorate,decoration=snake] (310:1.2) -- +(0:1.1) ;

\filldraw (0.65,1) rectangle (0.85,0.8);
\filldraw (180:0.75) circle (0.075);

\draw (0,-1.7) node [align=center] {(h)};

\node [left][ultra thick] at (-1.7,0) {$h$};
\node [above][ultra thick] at (0,0.8) {$\phi^{\pm}$};
\node [below][ultra thick] at (0,-0.8) {$W^{\pm}$};
\node [right][ultra thick] at (1,0) {$W^{\pm}$};
\node [right][ultra thick] at (1.8,0.9) {$\gamma$};
\node [right][ultra thick] at (1.8,-0.95) {$\gamma$};

\end{tikzpicture}
\begin{tikzpicture}

\draw  [dashed] [thick] (-1.7,0) -- (-0.75,0);

\draw[decorate,decoration=snake] (180:0.75) -- +(30:1.75) ;
\draw  [->][ultra thick]  (70:0.55)  -- + (30:0.01) ;

 \draw[decorate,decoration=snake] (180:0.75) -- +(330:1.75) ;
\draw  [->][ultra thick]  (290:0.55)  -- + (330:-0.01) ;

\draw[decorate,decoration=snake] (50:1.2) -- +(270:1.75) ;
\draw  [->][ultra thick]  (0:0.8)  -- + (270:0.01) ;

\draw[decorate,decoration=snake] (50:1.2) -- +(0:1.1) ;
\draw[decorate,decoration=snake] (310:1.2) -- +(0:1.1) ;

\filldraw (-0.6,-0.1) rectangle (-0.8,0.1);

\draw (0,-1.7) node [align=center] {(i)};

\node [left][ultra thick] at (-1.7,0) {$h$};
\node [above][ultra thick] at (0,0.8) {$W^{\pm}$};
\node [below][ultra thick] at (0,-0.8) {$W^{\pm}$};
\node [right][ultra thick] at (1,0) {$W^{\pm}$};
\node [right][ultra thick] at (1.8,0.9) {$\gamma$};
\node [right][ultra thick] at (1.8,-0.95) {$\gamma$};

\end{tikzpicture}
\hspace{5cm}
\begin{tikzpicture}
\end{tikzpicture}
\hspace{5.2cm}
\begin{tikzpicture}
\filldraw (0.6,-0.1) rectangle (0.8,0.1);

\draw  [dashed] [thick] (-1.2,0) -- (0.6,0);


\draw[decorate,decoration=snake] (0:0.75) -- +(30:1.30) ;
\draw[decorate,decoration=snake] (0:0.75) -- +(330:1.30) ;

\node [left][ultra thick] at (-1.3,0) {$h$};
\node [right][ultra thick] at (1.8,0.9) {$\gamma$};
\node [right][ultra thick] at (1.8,-0.95) {$\gamma$};

\draw (0.5,-1.5) node [align=center] {(j)};

\end{tikzpicture}

\caption{One loop diagrams contributing to $H \rightarrow \gamma \gamma$ decay through interactions in $\mathcal{L}_{eff}(v^0,v)$. Arrows on propagators indicate charge flow. The insertion
of the Effective Lagrangian in the diagram is indicated with a black square. Diagrams (f-i) have mirror diagrams that are not shown, where the photons
are exchanged in a less trivial manner than in diagrams (a-e). Diagram (j) corresponds to the insertion of the one loop counterterms present in $\mathcal{L}_{eff}$. Here, $h$ is the Higgs field, $\phi_{0,\pm}$ are the Goldstone bosons and $W$, $Z$ and $\gamma$ are the gauge fields.}
\label{Fig:1}

\end{figure}
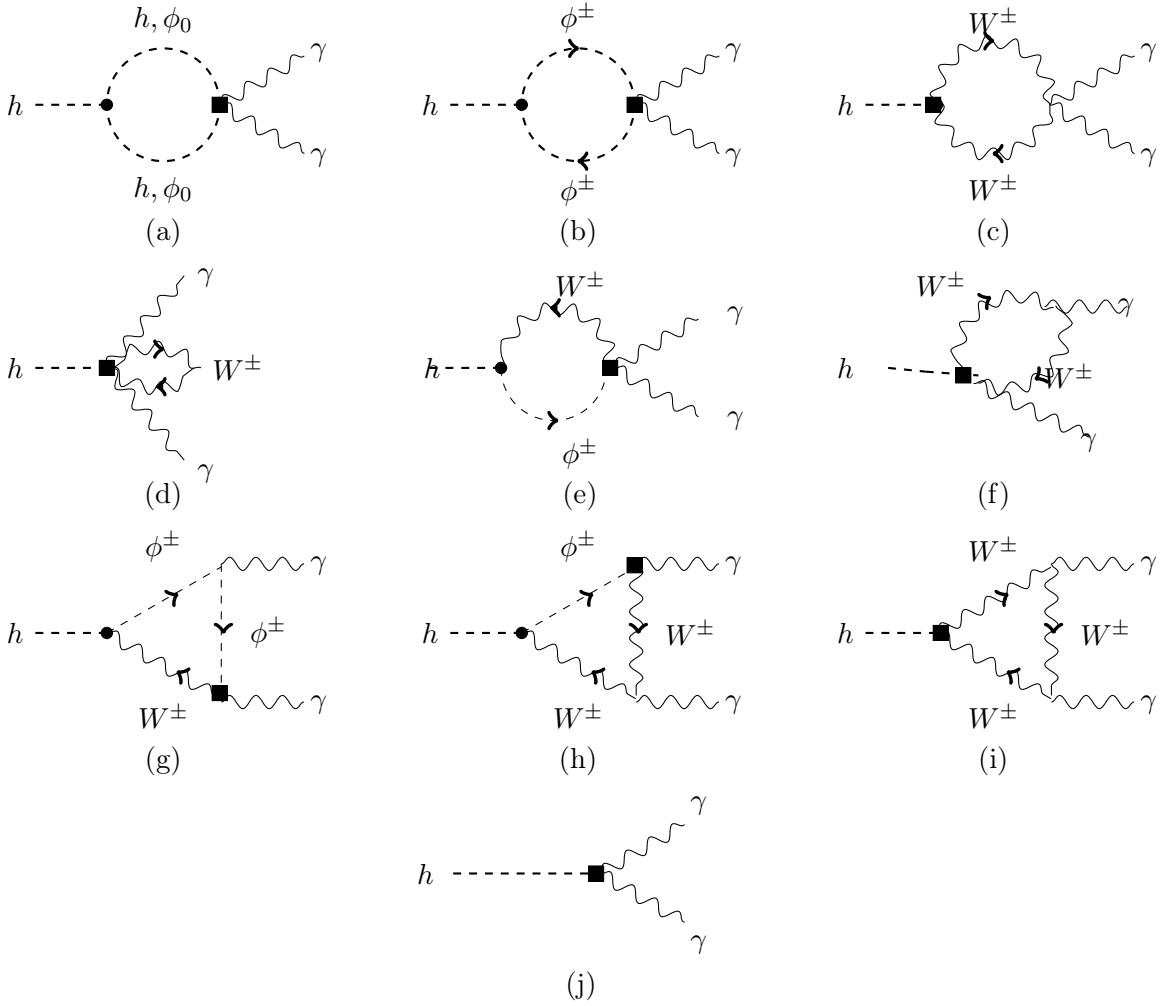

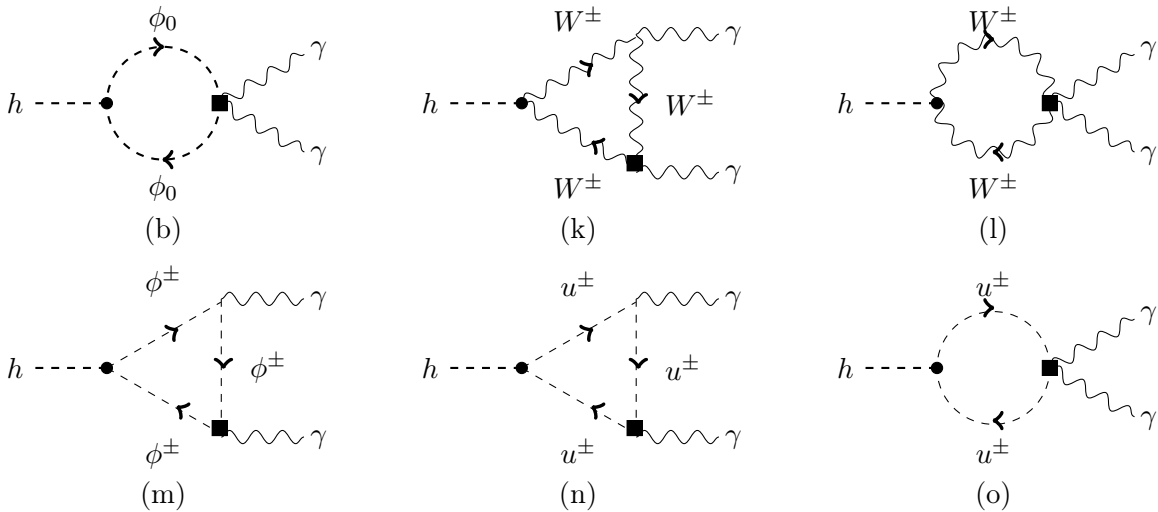
\begin{figure}[!h]
\begin{tikzpicture}[
decoration={
	markings,
	mark=at position 0.55 with {\arrow[scale=1.5]{stealth'}};
}]

\node [left][ultra thick] at (-1.7,0) {$h$};
\node [above][ultra thick] at (0,0.8) {$\phi_0$};
\node [below][ultra thick] at (0,-0.8) {$\phi_0$};
\node [right][ultra thick] at (1.80,0.7) {$\gamma$};
\node [right][ultra thick] at (1.80,-0.7) {$\gamma$};

\draw [thick] [dashed] (0,0) circle (0.75);

\filldraw (0.65,-0.1) rectangle (0.85,0.1);

\filldraw (180:0.75) circle (0.075);

\draw[decorate,decoration=snake] (0:0.75) -- +(30:1.30) ;
\draw[decorate,decoration=snake] (0:0.75) -- +(330:1.30) ;
\draw  [->][ultra thick]  (90:0.75)  -- + (0:0.01) ;
\draw  [->][ultra thick]  (270:0.75)  -- + (0:-0.01) ;


\draw  [dashed] [thick] (-1.7,0) -- (-0.75,0);

\draw (0,-1.7) node [align=center] {(b)};

\end{tikzpicture}
\hspace{0.8cm}
\begin{tikzpicture}

\draw  [dashed] [thick] (-1.7,0) -- (-0.75,0);

\draw[decorate,decoration=snake] (180:0.75) -- +(30:1.75) ;
\draw  [->][ultra thick]  (70:0.55)  -- + (30:0.01) ;

 \draw[decorate,decoration=snake] (180:0.75) -- +(330:1.75) ;
\draw  [->][ultra thick]  (290:0.55)  -- + (330:-0.01) ;

\draw[decorate,decoration=snake] (50:1.2) -- +(270:1.75) ;
\draw  [->][ultra thick]  (0:0.8)  -- + (270:0.01) ;

\draw[decorate,decoration=snake] (50:1.2) -- +(0:1.1) ;
\draw[decorate,decoration=snake] (310:1.2) -- +(0:1.1) ;

\filldraw (0.65,-0.9) rectangle (0.85,-0.7);

\filldraw (180:0.75) circle (0.075);
\draw (0,-1.7) node [align=center] {(k)};

\node [left][ultra thick] at (-1.7,0) {$h$};
\node [above][ultra thick] at (0,0.8) {$W^{\pm}$};
\node [below][ultra thick] at (0,-0.8) {$W^{\pm}$};
\node [right][ultra thick] at (1,0) {$W^{\pm}$};
\node [right][ultra thick] at (1.8,0.9) {$\gamma$};
\node [right][ultra thick] at (1.8,-0.95) {$\gamma$};

\end{tikzpicture}
\hspace{0.8cm}
\begin{tikzpicture}

\draw  [decorate,decoration=snake] (0,0) circle (0.75);

\filldraw (0.65,-0.1) rectangle (0.85,0.1);

\filldraw (180:0.75) circle (0.075);

\draw[decorate,decoration=snake] (0:0.75) -- +(30:1.30) ;
\draw[decorate,decoration=snake] (0:0.75) -- +(330:1.30) ;
\draw  [->][ultra thick]  (90:0.85)  -- + (0:0.01) ;
\draw  [->][ultra thick]  (270:0.65)  -- + (0:-0.01) ;


\draw  [dashed] [thick] (-1.7,0) -- (-0.75,0);

\draw (0,-1.7) node [align=center] {(l)};

\node [left][ultra thick] at (-1.7,0) {$h$};
\node [above][ultra thick] at (0,0.8) {$W^{\pm}$};
\node [below][ultra thick] at (0,-0.8) {$W^{\pm}$};
\node [right][ultra thick] at (1.80,0.7) {$\gamma$};
\node [right][ultra thick] at (1.80,-0.7) {$\gamma$};

\end{tikzpicture}

\begin{tikzpicture}

\draw  [dashed] [thick] (-1.7,0) -- (-0.75,0);

\draw[dashed] (180:0.75) -- +(30:1.75) ;
\draw  [->][ultra thick]  (70:0.55)  -- + (30:0.01) ;

 \draw[dashed] (180:0.75) -- +(330:1.75) ;
\draw  [->][ultra thick]  (290:0.55)  -- + (330:-0.01) ;

\draw[dashed] (50:1.2) -- +(270:1.75) ;
\draw  [->][ultra thick]  (0:0.8)  -- + (270:0.01) ;

\draw[decorate,decoration=snake] (50:1.2) -- +(0:1.1) ;
\draw[decorate,decoration=snake] (310:1.2) -- +(0:1.1) ;

\filldraw (0.65,-0.9) rectangle (0.85,-0.7);

\filldraw (180:0.75) circle (0.075);
\draw (0,-1.7) node [align=center] {(m)};

\node [left][ultra thick] at (-1.7,0) {$h$};
\node [above][ultra thick] at (0,0.8) {$\phi^{\pm}$};
\node [below][ultra thick] at (0,-0.8) {$\phi^{\pm}$};
\node [right][ultra thick] at (1,0) {$\phi^{\pm}$};
\node [right][ultra thick] at (1.8,0.9) {$\gamma$};
\node [right][ultra thick] at (1.8,-0.95) {$\gamma$};

\end{tikzpicture}
\hspace{0.8cm}
\begin{tikzpicture}

\draw  [dashed] [thick] (-1.7,0) -- (-0.75,0);

\draw[dashed] (180:0.75) -- +(30:1.75) ;
\draw  [->][ultra thick]  (70:0.55)  -- + (30:0.01) ;

 \draw[dashed] (180:0.75) -- +(330:1.75) ;
\draw  [->][ultra thick]  (290:0.55)  -- + (330:-0.01) ;

\draw[dashed] (50:1.2) -- +(270:1.75) ;
\draw  [->][ultra thick]  (0:0.8)  -- + (270:0.01) ;

\draw[decorate,decoration=snake] (50:1.2) -- +(0:1.1) ;
\draw[decorate,decoration=snake] (310:1.2) -- +(0:1.1) ;

\filldraw (0.65,-0.9) rectangle (0.85,-0.7);
\filldraw (180:0.75) circle (0.075);

\draw (0,-1.7) node [align=center] {(n)};

\node [left][ultra thick] at (-1.7,0) {$h$};
\node [above][ultra thick] at (0,0.8) {$u^{\pm}$};
\node [below][ultra thick] at (0,-0.8) {$u^{\pm}$};
\node [right][ultra thick] at (1,0) {$u^{\pm}$};
\node [right][ultra thick] at (1.8,0.9) {$\gamma$};
\node [right][ultra thick] at (1.8,-0.95) {$\gamma$};

\end{tikzpicture}
\hspace{0.8cm}
\begin{tikzpicture}

\draw  [dashed] (0,0) circle (0.75);

\filldraw (0.65,-0.1) rectangle (0.85,0.1);

\filldraw (180:0.75) circle (0.075);

\draw[decorate,decoration=snake] (0:0.75) -- +(30:1.30) ;
\draw[decorate,decoration=snake] (0:0.75) -- +(330:1.30) ;
\draw  [->][ultra thick]  (90:0.8)  -- + (0:0.01) ;
\draw  [->][ultra thick]  (270:0.7)  -- + (0:-0.01) ;


\draw  [dashed] [thick] (-1.7,0) -- (-0.75,0);

\draw (0,-1.7) node [align=center] {(o)};

\node [left][ultra thick] at (-1.7,0) {$h$};
\node [above][ultra thick] at (0,0.8) {$u^{\pm}$};
\node [below][ultra thick] at (0,-0.8) {$u^{\pm}$};
\node [right][ultra thick] at (1.80,0.7) {$\gamma$};
\node [right][ultra thick] at (1.80,-0.7) {$\gamma$};

\end{tikzpicture}

\caption{Diagrams contributing to $H \rightarrow \gamma \gamma$ decay due to $\mathcal{L}_{eff}(v^2)$. The divergent terms exactly cancel in this class of contributions, as expected. $u^{\pm}$ are ghost fields.}
\label{Fig:3}

\end{figure}

In this section, we explicitly demonstrate that the counterterm subtractions render the theory finite and cancel the $\xi$ gauge
dependence of the divergent terms while doing so. The required one loop diagrams explicitly calculated are shown in Figs.~\ref{Fig:1},~\ref{Fig:3}. To calculate the diagrams in the BFM, the Feynman rules of the theory, determined using the results of Ref.~\cite{Denner:1994xt}, are used. Our gauge coupling conventions differ from those chosen in Ref.~\cite{Denner:1994xt}, so we summarize our conventions in Appendix \ref{appendix}.
It is convenient to introduce the notation
\bea
 A_{\alpha \beta}^{h\gamma\gamma} &=&  \langle h | h \, A^{\mu \, \nu} \, A_{\mu \, \nu}| \gamma(p_a, \alpha),\gamma(p_b, \beta) \rangle
= - 4 \, \left(p_a \cdot p_b \, g^{\alpha \, \beta} - p_a^\beta \, p_b^\alpha \right), \nn \\
\mathcal{C}_\epsilon &=& \frac{i \, e^2 \, v}{16 \, \pi^2 \, \epsilon},  \quad \quad \mathcal{C}_\epsilon^{HW} = \frac{4 \, \mathcal{C}_\epsilon \, g_2^2 \, C_{HW}}{3},
\eea
to simplify results. The divergent contributions of the diagrams shown in Fig.~\ref{Fig:1} are as follows

\bea
i \mathcal{A}_a &=&  \mathcal{C}_\epsilon  \, C_{\gamma \, \gamma} \, (- 4 \, \lambda - \frac{1}{4}\,  (g_1^2+g_2^2)\, \xi) \, A_{\alpha \beta}^{h\gamma\gamma},  \\
i \mathcal{A}_b &=&  \mathcal{C}_\epsilon  (C_{\gamma \, \gamma} + 2 \, C_{HWB}) \, (- 2 \, \lambda - \frac{1}{2}\, g_2^2\, \xi) \, A_{\alpha \beta}^{h\gamma\gamma}, \\
i \mathcal{A}_c &=&   \mathcal{C}_\epsilon  g_2^2 \, C_{HW} \left[\left(1 - \frac{1}{\xi} \right) \frac{4}{3} \, p_a^\beta \, p_b^\alpha - 12 \, m_W^2 (3  + \frac{1}{\xi})\, g^{\alpha \, \beta}
+  \left(13 + \frac{5}{\xi} \right) \, \frac{4 \, p_a \cdot p_b}{3} \, g^{\alpha \, \beta} \right],  \\
i \mathcal{A}_d &=&   6 \, \mathcal{C}_\epsilon \, g_2^2 \, C_{HW}  \, m_W^2 \, \left[3 +  \, \xi^2 \right] \, g^{\alpha \, \beta},  \\
i \mathcal{A}_e &=&  \mathcal{C}_\epsilon \, g_2^2 \,  C_{HWB} \, \frac{(3 + \xi)}{2} \,  A_{\alpha \beta}^{h\gamma\gamma}, \\
i \mathcal{A}_f &=& 4 \, g_2^2 \, \mathcal{C}_\epsilon
\left[- 3 m_W^2 (\xi^2 + 3) \, C_{HW}  \, g^{\alpha \, \beta} - \frac{(\xi + 3)}{4} (2 C_{HW} - C_{HWB} ) A_{\alpha \beta}^{h\gamma\gamma} \right],  \\
i \mathcal{A}_g &=&  \mathcal{C}_\epsilon \, g_2^2 \, C_{HWB} \, (-1) A_{\alpha \beta}^{h\gamma\gamma},  \\
i \mathcal{A}_h &=& \mathcal{C}_\epsilon \, g_2^2 \,  C_{HWB} \, \frac{- \left(1+\xi \right)}{2} \, A_{\alpha \beta}^{h\gamma\gamma}, \\
i \mathcal{A}_i &=& \mathcal{C}_\epsilon^{HW}  \!  \left[\! \left(6 \, \xi + 17 + \frac{1}{\xi} \right)
p_a^\beta p_b^\alpha + \frac{9}{2} m_W^2 (\xi^2 + 9+ \frac{2}{\xi} ) g^{\alpha \, \beta}
-  \left(6 \, \xi + 31 + \frac{5}{\xi} \right) g^{\alpha \beta} p_a \cdot p_b \right]. \nn \\
\eea
The $\xi$ gauge dependence of the $Z_h$ and vev counterterms cancel against the sum of $i \mathcal{A}_{a,b}$.
The one loop counterterms of the operators in Eqn.~\ref{countertermmatrix} are gauge independent. As a result all of the remaining $\xi$ dependence in $i \mathcal{A}_{a-j}$ must cancel.
This occurs in a nontrivial manner. The sum of the divergent results of $i \mathcal{A}_{a-i}$ is
\bea
i \mathcal{A}_{a..i} = \frac{\mathcal{C}_\epsilon}{2}  \, \left[C_{HWB} (6 \, g_2^2 + 4 \, \lambda) - 12 \, \lambda  (C_{HB} + C_{HW})\right] \, A_{\alpha \beta}^{h\gamma\gamma}.
\eea
Here we are suppressing the dependence on the gauge parameter that exactly cancels against the gauge dependence in the $ \sqrt{Z_h},\sqrt{Z_v} + \delta v/v$ counterterms
in  $\mathcal{A}_{j}$.
The gauge independent divergent contributions exactly cancel the insertion of the gauge independent counterterms for the $O_i$ in $\mathcal{A}_{j}$, which gives
\bea
\langle h | h \,  \sqrt{Z_h} \, \left( v  \sqrt{Z_v} + \delta v \right) \, \left(\mathcal{N}_{HB}  + \mathcal{N}_{HW} - \mathcal{N}_{HWB} \right) e^2 \, A_{\mu\nu}A^{\mu\nu} | \gamma(p_a, \alpha) \, \gamma(p_b, \beta) \rangle, \nn \\
 =  -\frac{\mathcal{C}_\epsilon}{2} \, \left[C_{HWB} (6 \, g_2^2 + 4 \, \lambda) - 12 \, \lambda  (C_{HB} + C_{HW})\right] \, A_{\alpha \beta}^{h\gamma\gamma},
\eea
establishing that the process is rendered finite as expected. This is another non-trivial check of the mixing results reported in Ref.~\cite{Grojean:2013kd}.
The remaining divergences from $i \mathcal{A}_{b, k-o}$ in Fig.~\ref{Fig:3} due to $\mathcal{L}_{eff}(v^2)$ must vanish as argued, which occurs via the intermediate results
\bea
i \mathcal{A}_b &=&  - i \mathcal{A}_m = \mathcal{C}_\epsilon  \, e^2 \, v^2 \, C_{HWB} \,  (8 \, \lambda + 2 \ g_2^2 \, \xi) \, g^{\alpha \, \beta},  \\
i \mathcal{A}_k &=& -i \mathcal{A}_{l} =  - \frac{1}{2} \, \mathcal{C}_\epsilon \, e^2 \, v^2  \, g_2^2 \, C_{HWB} \, \left( 9+ \frac{3}{\xi} + \xi + 3 \, \xi^2 \right) \, g^{\alpha\beta}, \\
i \mathcal{A}_n &=& - i \mathcal{A}_o  = \mathcal{C}_\epsilon \, e^2 \, v^2  \, g_2^2 \, C_{HWB} \,4 \, \xi \, g^{\alpha\beta}.
\eea

\section{Finite terms}\label{finite}

The finite terms in the calculation come about from expanding the results of the diagrams in Figs.~\ref{Fig:1},\ref{Fig:3} to
$\mathcal{O}(\epsilon^0)$, as well as from finite one loop terms defined via renormalization conditions in the  $\rm \overline{MS}$ scheme for the fields and couplings entering at tree level in the
calculation.  The latter correspond to the vacuum expectation value ($\delta v$),  the renormalization conditions fixing the external two point functions for the Higgs ($\delta R_h$) and the photon  ($\delta R_A$) fields,
and the definition of the electric coupling $e$, which fixes $\delta R_e$.\footnote{The definition of these so-called R factors is well explained in Refs.~\cite{Denner:1991kt,Chiu:2009mg}.} These  BSM physics contributions enter the $S$ matrix element as
\bea\label{finite3}
\langle h(p_h) |S| \gamma(p_a, \alpha),\gamma(p_b, \beta) \rangle_{BSM} =(1 + \frac{\delta R_{h}}{2}) \, (1 + \delta R_{A})\, (1+ \delta R_e)^2 \, i \, \sum_{x = a..o} \, \mathcal{A}_{x}.
\eea
Note that $\delta v$ appears explicitly in $\mathcal{A}_{j}$.
We choose to use a scheme~\cite{Denner:1994xt,Denner:1991kt} that utilizes the BFM, where the finite on shell renormalization conditions are defined as follows:\footnote{In calculating the renormalization conditions with the BFM,  the two and three point functions and all external fields are treated as classical. An alternative scheme where systematically the two and three point functions are treated as
having mixed classical and quantum external states, combined with a suitable modification of the results due to Figs.~\ref{Fig:1},\ref{Fig:3}, can also be consistent.}

\begin{itemize}
\item The tree level vacuum expectation value of the Higgs is defined by the potential
\bea
\mathcal{L}_V = - \lambda \left(H^\dagger \, H - \frac{v^2}{2} \right)^2.
\eea
The one loop correction ($\delta v$) to the vacuum expectation value is fixed
by the condition that the one point function of the Higgs field vanishes to one loop order, including $\delta v$ in the definition of $H$ in Eqn.~\ref{Hexpansion}.
Including one loop corrections in the BFM, shown in Fig.~\ref{tadpoles2},
the finite terms linear in the Higgs field are modified to
\bea
T &=&   \, m_h^2 \, h \,v \, \frac{1}{16\pi^2}  \left[- 16 \pi^2 \, \frac{\delta v}{v}   + 3  \, \lambda \left(1+ \log \left[\frac{\mu^2}{m_h^2} \right] \right) + \frac{1}{4} \, g_2^2 \, \xi   \left(1+ \log \left[\frac{\mu^2}{\xi \, m_W^2} \right] \right)  \right.,  \\
&\,& \hspace{2.3cm} \left. + \frac{1}{8}(g_1^2+g_2^2) \, \xi \, \left(1+ \log \left[\frac{\mu^2}{\xi \, m_Z^2} \right] \right)
- 2 \sum_f y_f^2  \, N_c \frac{m_f^2}{m_h^2}\left(1+ \log \left[\frac{\mu^2}{m_f^2} \right] \right)  \right., \nn \\
&\,&  \hspace{2.3cm} \left.  + \frac{g_2^2}{2} \frac{m_W^2}{m_h^2} \left(1+ 3\log \left[\frac{\mu^2}{ m_W^2} \right] \right)
 +\frac{1}{4} (g_1^2 + g_2^2) \frac{m_Z^2}{m_h^2} \left(1+ 3\log \left[\frac{\mu^2}{m_Z^2} \right] \right) \right]. \nn
\eea
Setting $T =0$ defines the vev at one loop and fixes the finite terms of $\delta v$. Gauge dependence is thus present in the one loop definition of the vacuum expectation value of the Higgs. The remaining contributions to $\Gamma(h \rightarrow \gamma \, \gamma)$ from the sum of all other "tadpole" diagrams and insertions of $\delta v$ vanish due to this chosen renormalization condition. These contributions are not shown in Figs.~\ref{Fig:1},\ref{Fig:3}.

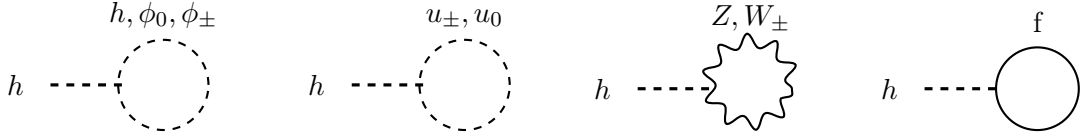
\begin{figure}

\begin{tikzpicture}

 \draw  [dashed]   [very thick](-0.5,0) -- (0.5,0);

\draw [thick] [dashed] (1,0) circle (0.60);

\node [above][ultra thick] at (1,0.6) {$h, \phi_0, \phi_{\pm}$};
\node [left][ultra thick] at (-0.7,0) {$h$};

\end{tikzpicture}
\hspace{0.7cm}
\begin{tikzpicture}

 \draw  [dashed]   [very thick](-0.5,0) -- (0.5,0);

\draw [thick] [dashed] (1,0) circle (0.60);

\node [above][ultra thick] at (1,0.6) {$u_{\pm},u_{0}$};
\node [left][ultra thick] at (-0.7,0) {$h$};

\end{tikzpicture}
\hspace{0.7cm}
\begin{tikzpicture}

 \draw  [dashed]   [very thick](-0.5,0) -- (0.5,0);

\draw [thick] [decorate,decoration=snake] (1,0) circle (0.55);

\node [above][ultra thick] at (1,0.6) {$Z,W_{\pm}$};
\node [left][ultra thick] at (-0.7,0) {$h$};

\end{tikzpicture}
\hspace{0.7cm}
\begin{tikzpicture}

 \draw  [dashed]   [very thick](-0.5,0) -- (0.5,0);

\draw [->][thick]  (1,0) circle (0.55);

\node [above][ultra thick] at (1,0.6) {f};
\node [left][ultra thick] at (-0.7,0) {$h$};

\end{tikzpicture}

\caption{One loop contributions to the one point expectation value $\langle h \rangle$. Here $u_{0, \pm}$ are ghost fields
and f is a fermion field. \label{tadpoles2}}
\end{figure}

\item The one loop correction to the  electric charge $e^{(0)} = (1+ \delta R_e) \, e^{(r)}$ is fixed by the on shell renormalization condition
defining the electric charge as the $e \, e \, \gamma$ coupling in the Thompson limit. Fixing all corrections to this vertex to vanish on shell
at zero momentum transfer fixes the finite terms given by $\delta R_e$. The Ward identities of the theory also fix
\bea\label{Ward}
\delta R_e = - \frac{1}{2} \delta R_{A},
\eea
where $\delta R_A$ is the wavefunction renormalization finite term fixed via the on shell renormalization condition of the photon field. Due to the above relation, no dependence on the photon two point function appears in the $S$ matrix element
at one loop.

\item The finite terms present in the wavefunction renormalization of the Higgs is defined via the on shell renormalization condition
\bea
\delta R_h =- \frac{\partial\Pi_{hh}(p^2)}{ \partial p^2}|_{p^2=m_h^2} .
\eea
where $p$ is the momentum of the Higgs field. Here $\Pi_{hh}(p^2)$ is the two point function of the Higgs calculated using the BFM, which is reported in Appendix \ref{App:finite}.
\end{itemize}
Combining these results gives the BSM contributions to the $S$ matrix element relevant for $\Gamma(h \rightarrow \gamma \, \gamma)$, due to the set of operators,
we have considered. Using the notation
\begin{align}
\mathcal{I}[m^2] & \equiv   \int_0^1 dx \,\log \left( \frac{m^2-m_h^2\, x \, (1-x)}{m_h^2} \right), &  \mathcal{I}_y[m^2] & \equiv \int_0^{1-x} dy \int_0^1 dx \, \frac{m^2}{m^2 - m_h^2\, x\, (1-x-y)},\nn
\end{align}
we find the result for this amplitude to be remarkably compact,\footnote{We have implemented Passarino-Veltman reductions~\cite{Passarino:1978jh} and made use of the tools Form, FormCalc and FeynCalc~\cite{Vermaseren:2000nd, Hahn:1998yk,Mertig:1990an} to carry out independent checks of the results.}
\bea\label{awesomeresult}
\frac{i \, \mathcal{A}_{total}^{NP}}{i \, v \, e^2\, A_{\alpha \beta}^{h\gamma\gamma} }&=& C_{\gamma \, \gamma} \left(1 + \frac{\delta R_{h}}{2} + \frac{\delta \, v}{v}\right), \nn \\
&+& \Bigg( \frac{C_{\gamma \, \gamma}}{16 \, \pi^2} \,  \left(\frac{g_1^2}{4}+\frac{3 \, g_2^2}{4}+ 6 \, \lambda \right)
+  \frac{C_{HWB}}{16 \, \pi^2} \,  \left(- 3 \, g_2^2 +4 \, \lambda  \right) \Bigg)  \, \log \left( \frac{m_h^2}{\Lambda^2}\right),
\nn \\
&+&\frac{C_{\gamma \, \gamma}}{16 \, \pi^2} \, \Bigg(
  \left(\frac{g_1^2}{4}+\frac{g_2^2}{4}+\lambda \right) \, \mathcal{I} [m_Z^2] + \left(\frac{g_2^2}{2}+2 \lambda
   \right) \, \mathcal{I} [m_W^2]+(\sqrt{3}  \, \pi  -6) \, \lambda \Bigg),
   \nn \\
   &+& \frac{C_{HWB}}{16 \, \pi^2} \,
   \Bigg( 2e^2\left( 1+ 6 \frac{m_W^2}{m_h^2} \right) - 2 \, g_2^2
   \, \left( 1+ \log
   \left( \frac{m_W^2}{m_h^2}\right)\right) + \left(4  \, \lambda -g_2^2 \right) \, \mathcal{I} [m_W^2],
   \nn \\
   &\,& \hspace{1.5cm} +\, 4 \left(3 e^2 -g_2^2 - 6e^2 \frac{m_W^2}{m_h^2}\right) \mathcal{I}_y[m_W^2] \Bigg), \nn \\
      &-&  \frac{ g_2^2 \, C_{HW}}{4 \, \pi^2} \,
   \Bigg( 3\, \frac{m_W^2}{m_h^2}
    + \left(4 - \frac{m_h^2}{m_W^2}  - 6 \frac{m_W^2}{m_h^2}\right) \, \mathcal{I}_y[m_W^2]  \Bigg).
\eea
This equation is the main result of this paper.
\subsection{Gauge independence of the results}
The results in Eqn.~\ref{awesomeresult} is gauge independent. Our use of the BFM allows
us to choose a gauge for the quantum fields (we use the convenient choice $\xi = 1$), while maintaining the
explicit gauge invariance of the result. We have also verified the gauge independence of the results with extensive checks of the calculation, which include
the following:
\begin{itemize}
\item{The gauge independence of the coefficient of $C_{\gamma \, \gamma}$ is due to a cancelation of the gauge dependence
in $\delta R_h$, $\delta v$ and the one loop contributions of diagrams a) and b) in Fig.~\ref{Fig:1} proportional to $C_{\gamma \, \gamma}$. We have explicitly verified this
gauge independence numerically using the CUBA library for numerical integration~\cite{Hahn:2004fe}.}
\item{The gauge independence of the remaining results proportional to $C_{HW}$ and $C_{HWB}$ in Eqn.~\ref{awesomeresult} due to Figs.~\ref{Fig:1},\ref{Fig:3} has also been checked numerically. Moreover, the results in Feynman $R_\xi$ gauge and unitary gauge have been compared analytically, finding exact agreement.\footnote{One aspect
of the gauge independence that is interesting is the generation of the $\lambda$ dependence for one loop corrections to $C_{HWB}$ in unitary gauge.
This dependence is due to diagram (b) in $R_\xi$ gauges, but is given by diagram (k) in unitary gauge, due to the appearance of a factor of $m_h^2$ in a numerator in the
one loop finite terms.}}
\end{itemize}
\subsection{Implications of the results for the one loop structure of the SMEFT}\label{implications}
The results in Eqn.~\ref{awesomeresult} are presented with terms grouped together in the following manner. The first line is the tree level result and the modification of this result due to the on shell renormalization conditions at one loop. The second line corresponds to log enhanced terms with $\mu \equiv \Lambda$. These results are consistent with the results reported in Refs.~\cite{Hagiwara:1993qt,Grojean:2013kd}.
However, the $\log(\mu)$ dependence proportional to $C_{\gamma \, \gamma}$ receives additional contributions from $R_{h}$ and $\delta v$. For $\xi = 1$, the relevant terms of this form are
\bea
16 \, \pi^2 \, (\frac{\delta v}{v})_{\log(\mu^2)} &=& \left(  3  \, \lambda
 + \frac{g_2^2}{4}  \,\Big( 1+ 6 \, \frac{m_w^2}{m_h^2}\Big) +\frac{g_1^2 + g_2^2}{8} \,\Big( 1 + 6 \, \frac{m_z^2}{m_h^2}\Big)   - 2 \,  \sum_f \, y_f^2 \,  \, N_c \,  \frac{m_f^2}{m_h^2}\right)  \log \left[\frac{\mu^2}{m_h^2} \right], \nn \\
16 \, \pi^2 \, (\delta R_h)_{\log(\mu^2)} &=& \left(g_1^2 + 3 g_2^2 - \sum_f y_f^2 N_c \right)  \log \left[\frac{\mu^2}{m_h^2} \right].
\eea
Here we have normalized all $\log(\mu^2)$ terms to the scale $m_h^2$. The scale dependence of the RG results in Ref.~\cite{Grojean:2013kd}
for $C_{\gamma \, \gamma}$ is obtained by adding
\bea
\Delta_{log(\mu)}^{h \, \gamma \, \gamma} = i \, v \, e^2 \, C_{\gamma \, \gamma} (\frac{\delta R_{h}}{2} - \frac{\delta \, v}{v}) \, A_{\alpha \beta}^{h\gamma\gamma},
\eea
to the result in Eqn.~\ref{awesomeresult}. The interpretation of this correction as a requirement to match the one loop scale dependence determined from the RG of the $O_i$, to the results
obtained in the full one loop result is fairly transparent. This correction is diagonal in the Wilson coefficient space of the $O_i$ at one loop. The existence of this correction nevertheless illustrates quite clearly the insufficiency of studying the one loop RG of the $O_i$ in the SMEFT as a direct proxy for the full one loop structure of the theory.

Furthermore, the last line of Eqn.~\ref{awesomeresult}
shows another important effect that is present in loop corrections and not captured in studying the structure of the RG of the $O_i$ alone.
Dependence on $C_{HW}$ is present at one loop, which is not absorbed into the dependence on $C_{\gamma \, \gamma}$ and is not hinted at in
the one loop structure of the RG. Nevertheless, such terms do exist. We refer to this class of terms as one loop "pure finite" terms, as no $\log(\mu)$ dependence is present for this class of contributions.

The form of the result in Eqn.~\ref{awesomeresult}
is exactly what one expects on general grounds. Although a particular combination of the operators $O_{HB},O_{HW}$ and $O_{HWB}$ corresponds to
$\Gamma(h \rightarrow \gamma \, \gamma)$ at tree level, there is no sense in which this particular combination of operators is preserved at one loop in the SMEFT.
The general expectation is that all three Wilson coefficients will appear as three independent parameters contributing to $\Gamma(h \rightarrow \gamma \, \gamma)$ at one loop.
This is exactly what is found in the explicit calculation of this process.

Finally, we note the effect of properly accounting for the class of effects due to $\mathcal{L}_{eff}(v^2)$ in the results. These contributions
do not contribute with divergences to be subtracted by the $O_i$ counterterms and so are expected to be pure finite terms. This can be understood due to the scaling in $v$
that these terms are proportional to. This is exactly what is found in the explicit calculation. Again, studies of only the $O_i$ RG of the SMEFT are blind to such contributions.
These contributions do not vanish at one loop when considering finite terms, and are the source of the $e^4$ terms in Eqn.~\ref{awesomeresult}.

For these reasons, {\it the structure of the RG is not a good proxy for the full one loop structure of the SMEFT.}

\section{Phenomenology}\label{phenomenology}

When considering the numerical effect of the Wilson coefficients of the $O_i$ on measured processes, there are two broad perspectives, one can adopt.
The SMEFT can be studied and experimentally constrained, while considering the SMEFT as a real and consistent theoretical formalism. This approach is a "bottom up"
EFT point of view, and does not impose a UV bias or prejudice as to the size of the Wilson coefficients of various operators.
Within this approach, consistent organizational schemes to power counting operators can be constructed. For example, a popular scheme is given by the Naive Dimensional Analysis (NDA) approach laid out in Ref.~\cite{Manohar:1983md}.
The NDA scheme is incomplete in some scenarios, but it can be consistently extended, see Refs.~\cite{Jenkins:2013sda,Buchalla:2013eza}.\footnote{One can also introduce consistent power counting schemes, with well defined assumptions, when one assumes that the underlying theory generating the $O_i$ is strongly interacting, see for example Ref.~\cite{Buchalla:2014eca}.} Within this bottom up EFT perspective, it is
interesting to note that  Ref.~\cite{Cheung:2015aba} has recently advanced an elegant operator level understanding of the approximate holomorphic structure of the SMEFT discussed in detail in
Ref.~\cite{Alonso:2014rga}. These works are concerned with statements on the SMEFT without invoking UV bias. The need to calculate finite terms
that need not have the structure of the RG, which is emphasized and explicitly demonstrated in this paper, is consistent with the discussions in Refs.~\cite{Alonso:2014rga,Cheung:2015aba}.

As an alternative to the bottom up approach, some literature employs UV dependent assumptions to classify operators. For some discussion along these lines see Refs.~\cite{Elias-Miro:2014eia,Elias-Miro:2013mua}.
We will not discuss UV bias in any great detail in this paper.
\subsection{Numerical results}
The perturbative results reported in this paper are one loop corrections to the effect of the chosen $O_i$ on the process $\Gamma(h \rightarrow \gamma \, \gamma)$.
The size of the $O_i$ Wilson coefficients is UV dependent. We treat the Wilson coefficients as free parameters, to be constrained by experiments,
multiplying the naive power counting suppression $\mathcal{O}(v^2/\Lambda^2)$. Currently the measured signal strength for $\gamma  \gamma$ is given by ATLAS~\cite{Aad:2014eha}  as
\begin{align}
\mu_{\gamma  \gamma} &= 1.17 \pm 0.27,
\end{align}
for $m_h = 125.4  \pm 0.4  \, {\rm GeV}$, while CMS reports~\cite{Khachatryan:2014ira}
\begin{align}
\mu_{\gamma  \gamma} &= 1.14^{+0.26}_{-0.23}\,,
\end{align}
for $m_h = 124.70 \pm 0.34  \, {\rm GeV}$.  The inferred results for the Higgs mass from these experiments are consistent~\cite{Aad:2015zhl}.
There is no evidence at this time for a deviation from the expectation in the SM.
The modified signal strength due to the chosen $O_i$ Wilson coefficients can be written as
\begin{align}
\label{hgamgammod}
\mu_{\gamma\gamma} \equiv {\Gamma(h \to \gamma \gamma) \over \Gamma^{\text{SM}}(h \to \gamma \gamma)} &\simeq \abs{1+ \frac{\mathcal{A}_{total}^{NP}}{A_{SM}}}^2.
\end{align}
The expression for $A_{SM}$ is well known from the literature, see for example Refs.~\cite{Ellis:1975ap,Bergstrom:1985hp,Manohar:2006gz}, and is given by
\bea
i \mathcal{A}_{SM} &=& \frac{i\, g \, e^2}{16 \, \pi^2 \, m_w} \int_0^1 dx \int_0^{1-x} dy  \\
&\,& \hspace{3cm} \, \Bigg( \frac{-4\, m_w^2 + 6 \, x\, y\, m_w^2 + x\ y\, m_h^2}{m_w^2 - x \, y \, m_h^2} +  \sum_f \, N_c\, Q_f^2 \, \frac{m_f^2 \, (1-4\, x\, y)}{m_f^2 - x\, y\, m_h^2} \Bigg) \, A_{\alpha \beta}^{h\gamma\gamma}. \nn
\eea
The corrections that we have computed are comparable in size to the RG log terms calculated in Refs.~\cite{Hagiwara:1993qt,Grojean:2013kd}. The results of the analyses of Refs.~\cite{Hagiwara:1993qt,Grojean:2013kd} are gauge independent. We compare our
results to Ref.~\cite{Grojean:2013kd}, which includes terms that were not calculated in Ref.~\cite{Hagiwara:1993qt}.
For the coefficient of $C_{HWB}$, the ratio of the full one loop terms,
to the results of Ref.~\cite{Grojean:2013kd} is given by
\bea
R_{CHWB} \simeq 1+ 0.7 \, \log^{-1} \frac{m_h^2}{\Lambda^2},
\eea
so that for $\rm {TeV}$ cut off scales, the RG log terms are larger by only a factor of $\sim 6$ (for $\Lambda \sim 1 \, {\rm TeV}$). For the coefficient of $C_{HW}$ not already included in $C_{\gamma \, \gamma}$, there are no RG terms. The pure finite term is given in Eqn.~\ref{awesomeresult}.
The ratio of the size of this term to the full one loop $C_{HWB}$ term, not already included in $C_{\gamma \, \gamma}$, is approximately
\bea
R_{CHWB/CHW} \simeq \frac{C_{HWB}}{C_{HW}} \left(0.5 + 0.7 \, \log{\frac{m_h^2}{\Lambda^2}}\right),
\eea
which makes clear that the pure finite terms are not negligible in favour of an RG analysis for cut off scales in the $\rm TeV$ range.
If perturbative corrections are incorporated in an analysis of future $\mu_{\gamma  \gamma}$ constraints, an RG analysis is simply insufficient.

Comparing the size of the one loop corrections included in our results that were neglected in past results,
it is clear that the appearance of the scale dependent term
\bea\label{biglog}
2 \sum_f y_f^2  \, N_c \frac{m_f^2}{m_h^2} \log \left[\frac{\mu^2}{m_h^2} \right],
\eea
in $\delta v$ is particularly problematic. For the top quark Yukawa, this is a dominant scale dependent log that is not captured in past RG analyses of the SMEFT. However, the interpretation of the numerical impact of this correction on $\mu_{\gamma \, \gamma}$ is more subtle.
Naively using $v \, + \,  \delta v$ as a numerical value of the vev extracted from a measured $G_F$ in $\mu^- \rightarrow e^- + \bar{\nu}_e + \nu_\mu$
would assign a gauge dependent quantity a numerical value. It is required that $\mu^- \rightarrow e^- + \bar{\nu}_e + \nu_\mu$  be calculated at one loop in the SMEFT\footnote{The required one loop calculation in the SM is known, see Ref.~\cite{Bardin:1999ak}.} to avoid this gauge dependence and allow the measurement to be used as a direct input into the prediction of $\mu_{\gamma \, \gamma}$.
This is beyond the scope of this work, but when doing so it is expected that the large log in Eqn.~\ref{biglog} will be absorbed in the numerical value of the vev in the SMEFT.

Finally, we note that there is some interest in the literature on the question of a one loop contribution from an operator that is not loop suppressed in  a UV matching, leading at one loop
to a contribution to $\mu_{\gamma \, \gamma}$ through the RG. This can occur in general, but does not occur in some cases if all UV physics is weakly coupled and renormalizable, see Refs.~\cite{Arzt:1994gp,Jenkins:2013fya,Manohar:2013rga,Henning:2014wua} for some coherent related discussion. Again we note that when considering the physical impact of such a scenario, studying the RG is only a part of a full one loop correction to a physical amplitude.

\section{Conclusions}\label{conclusion}
In this paper we have calculated one loop finite terms for the contribution
of the operators $O_{HB},O_{HWB}$ and $O_{HW}$ to the decay $\Gamma(h \rightarrow \gamma \, \gamma)$ in the SMEFT. The terms calculated are not small
in general compared to terms that have been calculated in past works using the RG of the SMEFT operators. If experimental bounds on $\Gamma(h \rightarrow \gamma \, \gamma)$
are to be studied to next to leading order in the SMEFT, these terms should not be neglected. In developing this result at one loop in the SMEFT, we
have also uncovered a number of interesting subtleties. These have broad implications when developing the SMEFT to next to leading order.

\section*{Acknowledgements}

We thank Poul Henrik Damgaard, Guido Festuccia, Alberto Guffanti, Gino Isidori, Simon Caron-Huot and Aneesh Manohar for helpful communication related to this work. In particular, we thank Simon and Aneesh, who also provided comments on the manuscript.
MT acknowledges generous support from the Villum Fonden and partial support by the Danish National Research Foundation (DNRF91).
The project leading to this application has received funding from the European Union's Horizon 2020 research
and innovation programme under the Marie Sklodowska-Curie grant agreement No 660876, HIGGS-BSM-EFT.
CH thanks Gudrun Heinrich and Sophia Borowka for their guidance in the acquaintance of the technical tool FormCalc, which was used to carry out calculations in this paper.

\appendix
\section{Conventions and Feynman Rules}\label{appendix}
We define our notational conventions and Feynman rules in this section. The covariant derivative sign convention is
defined as $D^\mu = \partial_\mu + i g_2 W^a_\mu T^a + i g_1 B_\mu Y$, with $Y$ the $U(1)$ Hypercharge generator.  Here $T^a = \sigma^a/2$, with $\sigma^a$ being the Pauli matrices.
The sign convention in the covariant derivative fixes the sign conventions in the  Yang-Mills part of the Lagrangian to be
\bea
W^a_{\mu \, \nu} &=& \partial_\mu W^a_\nu - \partial_\nu W^a_\mu - g_2 \, \epsilon_{abc} \, W^b_\mu \, W^c_\nu, \nn \\
D_\mu W^a_\nu &=& \partial_\mu W^a_\nu - g_2 \,\epsilon^{abc} W_{b,\mu} W_{c, \nu}.
\eea
The conventions used here are consistent with Refs.~\cite{Manohar:2000dt,Alonso:2013hga}. We use the BFM,
and the background fields are introduced with a hat superscript. Quantum fields are denoted without the hat superscript.
The gauge fixing is given by
\bea
\mathcal{L}_{GF} &=&  - \frac{1}{2 \, \xi_W} \sum_a \left[ \partial_\mu W^{a,\mu} -  g_2 \,\epsilon^{abc} \hat{W}_{b,\mu} W_{c}^{\mu} + i \, g_2 \, \frac{\xi}{2} \left(\hat{H}^\dagger_i \sigma^a_{ij} H_j - H^\dagger_i \sigma^a_{ij} \hat{H}_j  \right) \right]^2, \nn \\
&-&  \frac{1}{2 \, \xi_B} \left[ \partial_\mu B^{\mu} + i \, g_1 \, \frac{\xi}{2} \left(\hat{H}^\dagger_i H_i - H^\dagger_i \hat{H}_i  \right) \right]^2.
\eea
From the gauge fixing term, choosing $\xi_B = \xi_W$, one directly finds
\bea\label{GFresult}
\mathcal{L}_{GF} = - \frac{1}{2 \,\xi} \left[(G^A)^2 + (G^Z)^2  + 2 \, G^+ \, G^-\right],
\eea
where
\bea
G^A &=& \partial_\mu A^\mu + i e \left(\hat{W}^+_\mu W^-_\mu - W^+_\mu \hat{W}^-_\mu \right) + i e \,  \xi \left(\hat{\phi}^- \phi^+ - \hat{\phi}^+ \phi^- \right), \nn \\
G^Z &=& \partial_\mu Z^\mu + i e \frac{c_w}{s_w} \left(\hat{W}^+_\mu W^-_\mu - W^+_\mu \hat{W}^-_\mu \right) + i e \xi \frac{1}{2 c_w \, s_w}(c_w^2 - s_w^2) \left(\hat{\phi}^- \phi^+ - \hat{\phi}^+ \phi^- \right), \nn \\
&-& e \, \xi \frac{1}{2 \, c_w s_w}  \left(\hat{\phi}_0 \,h - \hat{h} \phi_0 - v \phi_0 \right), \nn \\
G^\pm &=& \partial^\mu W^\pm_\mu  \pm i \, e \, \left[\hat{A}^\mu  + \frac{c_w}{s_w} \, \hat{Z}^\mu \right] W^\pm_\mu \mp  i e \left(A^\mu +\frac{c_w}{s_w} Z^\mu \right) \hat{W}^\pm_\mu, \nn \\
&-& e \,  \xi \, \frac{1}{2 s_w} \, \left((v+ \hat{h} \mp i \hat{\phi}_0) \phi^\pm  - (h \mp i \phi_0) \hat{\phi}^\pm \right).
\eea
This result is consistent (up to sign conventions) with the result in Ref.~\cite{Denner:1994xt}. From this, the Feynman rules of the theory are straightforward
to derive.
\subsection{Digression on gauge fixing}
Gauge fixing in the SMEFT has some interesting subtleties compared to gauge fixing in the SM. Fundamentally, the number of degrees of freedom being fixed are the same, and in this sense standard t'Hooft $R_\xi$ gauge fixing~\cite{'tHooft:1971fh,'tHooft:1971fh}~can be imposed.
However, the relationship between the $\rm SU(2)_L \times U(1)_Y$ gauge fields $W$ and $B$ and the mass eigenstate fields is modified
in the SMEFT.  In this sense, it is clearly legitimate to consider more carefully the question -- What exactly is being gauged in the SMEFT?

The fact that the relation between the $W$ and $B$ fields and the physical propagating mass eigenstate fields changes order by order in the power counting of the theory,
as do the mixing angles, has a number of interesting consequences. These contributions can physically contribute to amplitudes in pure finite terms at one loop,
as we have shown. As such, the interesting consequences of gauge fixing in the SMEFT are worthy of some comment.

For example, the ghost Lagrangian can be derived from Eqn.~\ref{GFresult} adding a standard Faddeev-Popov term~\cite{Faddeev:1967fc}
to the Lagrangian, where
\bea
\mathcal{L}_{FP} = -\bar{u}^\alpha \, \frac{\delta G^\alpha}{\delta \theta^\beta} \, u^\beta.
\eea
Here $\alpha, \beta$ are summed over the physical mass eigenstate fields, and $\delta G^\alpha/\delta \theta^\beta$ is the variation
of the gauge fixing terms under the infinitesimal quantum gauge fixing transformations. The gauge transformations in Ref.~\cite{Denner:1994xt} can be used to derive
the ghost Lagrangian, including the effect of redefining the mixing angles and states with SMEFT corrections.
In this manner, at $\mathcal{O}(v^2/\Lambda^2)$, we observe that the Wilson coefficient $C_{HWB}$ sources ghost interactions. Another interesting consequence of gauge fixing in the SMEFT
is the presence of the interaction terms
\bea
 - \frac{c_w \, s_w}{ \xi_B \, \xi_W} (\xi_B - \xi_W) \left(\partial^\mu A_\mu \, \partial^\nu \, Z_\nu\right)
- \frac{C_{HWB} v^2 (s_w^2 -c_w^2) (s_w^2 \xi_B + c_w^2 \xi_W)}{\xi_B \, \xi_W}  \left(\partial^\mu A_\mu \, \partial^\nu \, Z_\nu\right)
 \cdots
\eea
where in this case the gauge fixing has not imposed the relation $\xi_B = \xi_W$. In the SM, the choice  $\xi_B = \xi_W$ is usually made in t'Hooft gauge fixing,
and as a consequence $A$-$Z$ mixing is not present at tree level. The same choice in the SMEFT results
in tree level $A$-$Z$ mixing. This is due to the redefinition of the mass eigenstate fields of the theory at $\mathcal{O}(v^2/\Lambda^2)$ in the SMEFT. These field redefinitions cause these terms to
result from the gauge fixing procedure.

The presence of such unanticipated interaction terms when generalising the SM into the SMEFT emphasizes the importance of not obscuring
the distinction between a basis of gauge invariant operators, and the
Effective Lagrangian.  The severe challenges of implementing a calculation of the form presented here, if the distinction is lost, should
be manifest.  This is a very serious limitation to the gauge dependent approach to the SMEFT discussed in
Refs.~\cite{Gupta:2014rxa,HiggsNonBasis,Falkowski:2015fla}, which is not well suited for long term use, as studies in the SMEFT are already being systematically
improved beyond tree level.

Conversely, avoiding some of the gauge dependence challenges in the SMEFT is a
feature in favour of the covariant derivative expansion discussed in Refs.~\cite{Gaillard:1985uh,Henning:2014wua,Drozd:2015kva} and by direct use of a well
defined non-redundant basis, such as the Warsaw basis of Ref.~\cite{Grzadkowski:2010es}.

\section{Higgs self energy}\label{App:finite}

The finite part of the Higgs self energy will define the R factors of the Higgs field and mass renormalization. In Feynman gauge, setting $\xi = 1$, the finite part of the Higgs self energy amplitude is given by

\begin{align}
 16 \, \pi^2 \, \mathcal{A}^{fin}_{a} & = \,  m_z^2\, \left( \frac{1}{4} \,( g_1^2+g_2^2) \left( 1 - \log
   \left(\frac{m_z^2}{\mu ^2}\right)\right)+\lambda \, \left(1-\log \left(\frac{m_z^2}{\mu
   ^2}\right)\right)\right)  + \, 3 \, \lambda \, m_h^2 \,\left(1- \log
   \left(\frac{m_h^2}{\mu ^2}\right)\right), \nn \\
& \, \, \, + m_w^2 \,\left(\frac{1}{2} \, g_2^2 \, \left(1- \log
   \left(\frac{m_w^2}{\mu ^2}\right)\right)+2\, \lambda \, \left(1- \log
   \left(\frac{m_w^2}{\mu ^2}\right)\right)\right),
   \nonumber \\
16 \, \pi^2 \, \mathcal{A}^{fin}_{b} &= \frac{1}{2} \, m_z^2 \, \left(g_1^2+g_2^2\right) \, \left(1-2 \, \log
   \left(\frac{m_z^2}{\mu ^2}\right)\right) + g_2^2 \, m_w^2 \, \left( 1 - 2 \, \log \left(\frac{m_w^2}{\mu^2}\right)\right),
\nonumber \\
16 \, \pi^2 \, \mathcal{A}^{fin}_{c} &  = \, - \frac{1}{2} \, m_z^2 \, \left(g_1^2+g_2^2\right)  \, \Bigg( 1-  \mathcal{I}[m_z^2] \Bigg) - g_2^2\, m_w^2 \, \Bigg(1 -  \mathcal{I}[m_w^2]\Bigg),\nn \\
16 \, \pi^2 \, \mathcal{A}^{fin}_d &= - 2  \,m_f^2 \, \sum_f \, y_f^2 \, N_c  \, \left(1 -  \log \left(\frac{m_f^2}{\mu ^2}\right) - 2 \, \mathcal{I}[m_f^2] \right)-p^2  \, \sum_f \, y_f^2 \, N_c \, \mathcal{I}[m_f^2], \nn \\
16 \, \pi^2 \,  \mathcal{A}^{fin}_{e} &  =  - \, 9  \, \lambda \,  m_h^2 \, \mathcal{I}[m_h^2] \, - \Bigg(m_h^2 \,  \left(g_2^2 + 2 \, \lambda\right)  + g_2^2  \, m_w^2 \Bigg) \, \mathcal{I}[m_w^2], \nn \\
  & \, \, \,  - \frac{1}{2} \, \Bigg(  m_h^2 \,
   \left(g_1^2+ g_2^2+ 2\lambda
   \right) +
   m_z^2 \, (g_1^2+g_2^2) \Bigg) \, \mathcal{I}[m_z^2]     , \nn \\
16 \, \pi^2 \,  \mathcal{A}^{fin}_{f} & = \, m_z^2 \, \left(g_1^2+g_2^2\right) \, \mathcal{I}[m_z^2] +  2  \, g_2^2 \,  m_w^2 \, \mathcal{I}[m_w^2], \nn \\
16 \, \pi^2 \,  \mathcal{A}^{fin}_g  &= -2 \, g_2^2 \, m_w^2 \, \Bigg( 1+ 2 \, \mathcal{I}[m_w^2]\Bigg)
 - m_z^2 \, \left(g_1^2 + g_2^2\right) \Bigg( 1+ 2 \, \mathcal{I}[m_z^2] \Bigg), \nn \\
16 \, \pi^2 \,  \mathcal{A}^{fin}_{h} & = \,  2  \, g_2^2 \, p^2 \, \mathcal{I}[m_w^2],  \nn \\
16 \, \pi^2 \,  \mathcal{A}^{fin}_i & =\,p^2 \, \left(g_1^2+g_2^2\right) \, \mathcal{I}[m_z^2],
  \end{align}
where $p$ is the momentum of the external Higgs fields. From the sum of these results, $R_h = 1+ \delta R_h$ can be found from,
\bea
\delta R_h =- \frac{\partial\Pi_{hh}(p^2)}{ \partial p^2}|_{p^2=m_h^2} .
\eea
For $\xi = 1$ we find the result for $16 \, \pi^2 \,  \delta R_h$,
\begin{align}
&   2 \,  \lambda \, \Big(6  - \sqrt{3} \pi - \mathcal{J}_x[m_z^2] - 2 \, \mathcal{J}_x[m_w^2] \Big) +\left(\sum_f \, y_f^2 \, N_c -g_1^2-3 g_2^2\right) \log
   \left(\frac{m_h^2}{\mu ^2}\right),
\nn \\
 & +2 \, g_2^2 \Bigg(
   \Big( \mathcal{J}_x[m_w^2]- \frac{1}{2} \Big)  \,  \left(1 -\frac{3 m_w^2}{m_h^2} \right)  - \mathcal{I}[m_w^2]\Bigg) +\left(g_1^2+g_2^2\right) \, \Bigg( \Big( \mathcal{J}_x[m_z^2] - \frac{1}{2}\Big)\left(1 - \frac{3
   m_z^2}{m_h^2}\right) -\mathcal{I}[m_z^2] \Bigg),
   \nn \\
   & +\sum_f \, y_f^2 \, N_c \Bigg(1+
   \left(1+\frac{2 m_f^2}{m_h^2}\right)\, \mathcal{I}[m_f^2] -\frac{2 m_f^2}{m_h^2} \,\log
   \left(\frac{m_f^2}{m_h^2}\right) \Bigg).
\end{align}
Here, we have used the notation
\bea
\mathcal{J}_{x}[m^2] & \equiv & \int_0^1 dx \, \frac{x \, m^2}{m^2 - m_h^2\, x\, (1-x)},
\eea
and we note that although we present this result with $\xi = 1$, it was determined in the background field method with
classical external Higgs fields assumed for the two point function.

\begin{figure}

\hspace{1.7cm}
\begin{tikzpicture}

 \draw  [dashed]   [very thick](-1,0) -- (1,0);

\draw [thick] [dashed] (0,0.5) circle (0.50);

\draw (0,-1) node [align=center] {(a)};

\node [above][ultra thick] at (0,1) {$h, \phi_0, \phi_{\pm}$};
\node [left][ultra thick] at (-0.3,-0.2) {$h$};

\end{tikzpicture}
\hspace{3.4cm}
\begin{tikzpicture}

\draw  [dashed]   [very thick](-1,0) -- (1,0);

\draw  [decorate,decoration=snake] (0,0.4) circle (0.45);

\draw (0,-1) node [align=center] {(b)};

\node [above][ultra thick] at (0,1) {$Z, W_{\pm}$};
\node [left][ultra thick] at (-0.3,-0.2) {$h$};

\end{tikzpicture}
\hspace{3.4cm}
\begin{tikzpicture}

 \draw  [dashed]   [very thick](-1,0) -- (1,0);

\draw [thick] [dashed] (0,0.5) circle (0.50);

\draw (0,-1) node [align=center] {(c)};

\node [above][ultra thick] at (0,1) {$u_0, u_{\pm}$};
\node [left][ultra thick] at (-0.3,-0.2) {$h$};

\end{tikzpicture}

\hspace{1.6cm}
\begin{tikzpicture}

\draw  [dashed]   [very thick](-1.1,0) -- (-0.5,0);
\draw  [dashed]   [very thick](0.5,0) -- (1.1,0);

\draw  (0,0) circle (0.50);

\draw (0,-1) node [align=center] {(d)};

\node [above][ultra thick] at (0,0.5) {$f$};
\node [left][ultra thick] at (-0.5,-0.2) {$h$};

\end{tikzpicture}
\hspace{3.2cm}
\begin{tikzpicture}

\draw  [dashed]   [very thick](-1.1,0) -- (-0.5,0);
\draw  [dashed]   [very thick](0.5,0) -- (1.1,0);

\draw [thick] [dashed] (0,0) circle (0.50);

\draw (0,-1) node [align=center] {(e)};

\node [above][ultra thick] at (0,0.5) {$h, \phi_0, \phi_{\pm}$};
\node [left][ultra thick] at (-0.5,-0.2) {$h$};

\end{tikzpicture}
\hspace{3.2cm}
\begin{tikzpicture}

\draw  [dashed]   [very thick](-1.1,0) -- (-0.5,0);
\draw  [dashed]   [very thick](0.5,0) -- (1.1,0);

\draw [thick] [dashed] (0,0) circle (0.50);

\draw (0,-1) node [align=center] {(f)};

\node [above][ultra thick] at (0,0.5) {$u_{0}, u_{\pm}$};
\node [left][ultra thick] at (-0.5,-0.3) {$h$};

\end{tikzpicture}

\hspace{1.6cm}
\begin{tikzpicture}

\draw  [dashed]   [very thick](-1.1,0) -- (-0.5,0);
\draw  [dashed]   [very thick](0.5,0) -- (1.1,0);

\draw [decorate,decoration=snake] (0,0) circle (0.45);

\draw (0,-1) node [align=center] {(g)};

\node [above][ultra thick] at (0,0.5) {$Z, W_{\pm}$};
\node [left][ultra thick] at (-0.5,-0.3) {$h$};

\end{tikzpicture}
\hspace{3.2cm}
\begin{tikzpicture}

\draw  [dashed]   [very thick](-1.1,0) -- (-0.45,0);
\draw  [dashed]   [very thick](0.5,0) -- (1.1,0);

\draw (0,-1) node [align=center] {(h)};

\node [above][ultra thick] at (0,0.5) {$W_{\pm}$};
\node [above][ultra thick] at (0,-0.5) {$\phi_{\mp}$};

\node [left][ultra thick] at (-0.5,-0.3) {$h$};

\draw [decorate,decoration=snake] (0.5,0) arc [radius=0.45, start angle=0, end angle= 190];

\draw [dashed] (-0.45,0) arc [radius=0.45, start angle=178, end angle= 360];

\end{tikzpicture}
\hspace{3.2cm}
\begin{tikzpicture}

\draw  [dashed]   [very thick](-1.1,0) -- (-0.45,0);
\draw  [dashed]   [very thick](0.5,0) -- (1.1,0);

\draw (0,-1) node [align=center] {(i)};

\node [above][ultra thick] at (0,0.5) {$Z$};
\node [above][ultra thick] at (0,-0.5) {$\phi_{0}$};

\node [left][ultra thick] at (-0.5,-0.3) {$h$};

\draw [decorate,decoration=snake] (0.5,0) arc [radius=0.45, start angle=0, end angle= 190];

\draw [dashed] (-0.45,0) arc [radius=0.45, start angle=178, end angle= 360];

\end{tikzpicture}

\label{ Higgsself}

\caption{Diagrams contributing to the Higgs self energy.}
\end{figure}
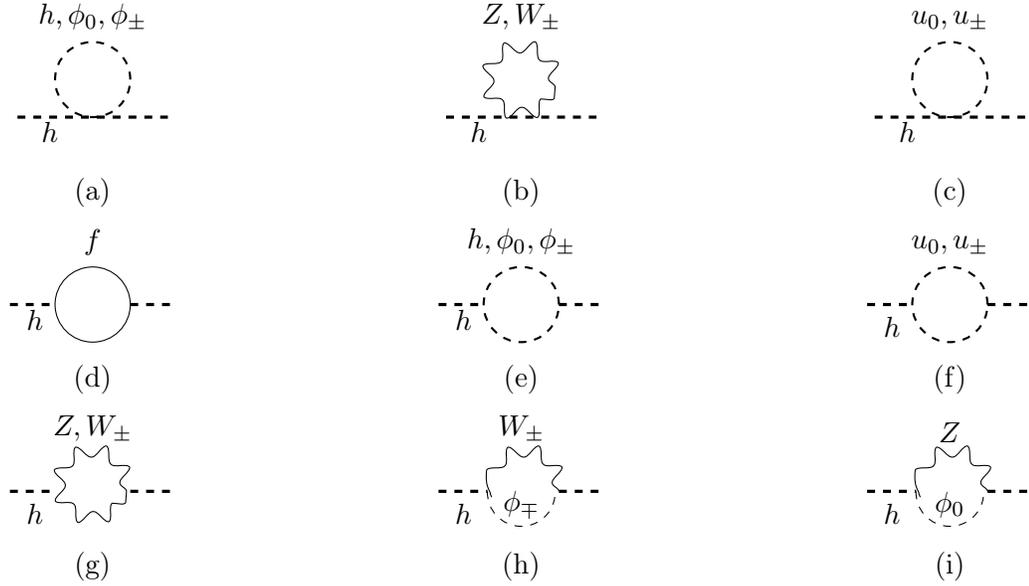

\bibliographystyle{JHEP}
\bibliography{RG}

\providecommand{\href}[2]{#2}\begingroup\raggedright\begin{thebibliography}{10}

\bibitem{Buchmuller:1985jz}
W.~Buchmuller and D.~Wyler, {\it {Effective Lagrangian Analysis of New
  Interactions and Flavor Conservation}},  {\em Nucl.Phys.} {\bf B268} (1986)
  621.

\bibitem{Grzadkowski:2010es}
B.~Grzadkowski, M.~Iskrzynski, M.~Misiak, and J.~Rosiek, {\it {Dimension-Six
  Terms in the Standard Model Lagrangian}},  {\em JHEP} {\bf 1010} (2010) 085,
  [\href{http://xxx.lanl.gov/abs/1008.4884}{{\tt arXiv:1008.4884}}].

\bibitem{Lehman:2014jma}
L.~Lehman, {\it {Extending the Standard Model Effective Field Theory with the
  Complete Set of Dimension-7 Operators}},  {\em Phys.Rev.} {\bf D90} (2014),
  no.~12 125023, [\href{http://xxx.lanl.gov/abs/1410.4193}{{\tt
  arXiv:1410.4193}}].

\bibitem{Grojean:2013kd}
C.~Grojean, E.~E. Jenkins, A.~V. Manohar, and M.~Trott, {\it {Renormalization
  Group Scaling of Higgs Operators and $\Gamma(h -> \gamma \gamma)$}},  {\em
  JHEP} {\bf 1304} (2013) 016, [\href{http://xxx.lanl.gov/abs/1301.2588}{{\tt
  arXiv:1301.2588}}].

\bibitem{Jenkins:2013zja}
E.~E. Jenkins, A.~V. Manohar, and M.~Trott, {\it {Renormalization Group
  Evolution of the Standard Model Dimension Six Operators I: Formalism and
  lambda Dependence}},  {\em JHEP} {\bf 1310} (2013) 087,
  [\href{http://xxx.lanl.gov/abs/1308.2627}{{\tt arXiv:1308.2627}}].

\bibitem{Jenkins:2013wua}
E.~E. Jenkins, A.~V. Manohar, and M.~Trott, {\it {Renormalization Group
  Evolution of the Standard Model Dimension Six Operators II: Yukawa
  Dependence}},  {\em JHEP} {\bf 1401} (2014) 035,
  [\href{http://xxx.lanl.gov/abs/1310.4838}{{\tt arXiv:1310.4838}}].

\bibitem{Alonso:2013hga}
R.~Alonso, E.~E. Jenkins, A.~V. Manohar, and M.~Trott, {\it {Renormalization
  Group Evolution of the Standard Model Dimension Six Operators III: Gauge
  Coupling Dependence and Phenomenology}},  {\em JHEP} {\bf 1404} (2014) 159,
  [\href{http://xxx.lanl.gov/abs/1312.2014}{{\tt arXiv:1312.2014}}].

\bibitem{Alonso:2014zka}
R.~Alonso, H.-M. Chang, E.~E. Jenkins, A.~V. Manohar, and B.~Shotwell, {\it
  {Renormalization group evolution of dimension-six baryon number violating
  operators}},  {\em Phys.Lett.} {\bf B734} (2014) 302,
  [\href{http://xxx.lanl.gov/abs/1405.0486}{{\tt arXiv:1405.0486}}].

\bibitem{Passarino:2012cb}
G.~Passarino, {\it {NLO Inspired Effective Lagrangians for Higgs Physics}},
  {\em Nucl.Phys.} {\bf B868} (2013) 416--458,
  [\href{http://xxx.lanl.gov/abs/1209.5538}{{\tt arXiv:1209.5538}}].

\bibitem{Chen:2013kfa}
C.-Y. Chen, S.~Dawson, and C.~Zhang, {\it {Electroweak Effective Operators and
  Higgs Physics}},  {\em Phys.Rev.} {\bf D89} (2014), no.~1 015016,
  [\href{http://xxx.lanl.gov/abs/1311.3107}{{\tt arXiv:1311.3107}}].

\bibitem{Henning:2014wua}
B.~Henning, X.~Lu, and H.~Murayama, {\it {How to use the Standard Model
  effective field theory}},  \href{http://xxx.lanl.gov/abs/1412.1837}{{\tt
  arXiv:1412.1837}}.

\bibitem{Englert:2014cva}
C.~Englert and M.~Spannowsky, {\it {Effective Theories and Measurements at
  Colliders}},  {\em Phys.Lett.} {\bf B740} (2015) 8--15,
  [\href{http://xxx.lanl.gov/abs/1408.5147}{{\tt arXiv:1408.5147}}].

\bibitem{Zhang:2014lya}
C.~Zhang, {\it {Effective approach to top-quark decay and FCNC processes at NLO
  accuracy}},  {\em J.Phys.Conf.Ser.} {\bf 556} (2014), no.~1 012030,
  [\href{http://xxx.lanl.gov/abs/1410.2825}{{\tt arXiv:1410.2825}}].

\bibitem{Grober:2015cwa}
R.~Grober, M.~Muhlleitner, M.~Spira, and J.~Streicher, {\it {NLO QCD
  Corrections to Higgs Pair Production including Dimension-6 Operators}},
  \href{http://xxx.lanl.gov/abs/1504.0657}{{\tt arXiv:1504.0657}}.

\bibitem{sing1}
G.~M. Pruna and A.~Signer, {\it {The $\mu\to e\gamma$ decay in a systematic
  effective field theory approach with dimension 6 operators}},  {\em JHEP}
  {\bf 1410} (2014) 14, [\href{http://xxx.lanl.gov/abs/1408.3565}{{\tt
  arXiv:1408.3565}}].

\bibitem{'tHooft:1975vy}
G.~'t~Hooft, {\it {The Background Field Method in Gauge Field Theories,
  Proceedings, Acta Universitatis Wratislaviensis No.368, Vol.1*, Wroclaw 1976,
  345-369}}, .

\bibitem{DeWitt:1967ub}
B.~S. DeWitt, {\it {Quantum Theory of Gravity. 2. The Manifestly Covariant
  Theory}},  {\em Phys.Rev.} {\bf 162} (1967) 1195--1239.

\bibitem{Abbott:1981ke}
L.~Abbott, {\it {Introduction to the Background Field Method}},  {\em Acta
  Phys.Polon.} {\bf B13} (1982) 33.

\bibitem{Denner:1994xt}
A.~Denner, G.~Weiglein, and S.~Dittmaier, {\it {Application of the background
  field method to the electroweak standard model}},  {\em Nucl.Phys.} {\bf
  B440} (1995) 95--128, [\href{http://xxx.lanl.gov/abs/hep-ph/9410338}{{\tt
  hep-ph/9410338}}].

\bibitem{Einhorn:1988tc}
M.~B. Einhorn and J.~Wudka, {\it {Screening of Heavy Higgs Radiative Effects}},
   {\em Phys.Rev.} {\bf D39} (1989) 2758.

\bibitem{Manohar:2000dt}
A.~V. Manohar and M.~B. Wise, {\it {Heavy quark physics}},  {\em
  Camb.Monogr.Part.Phys.Nucl.Phys.Cosmol.} {\bf 10} (2000) 1--191.

\bibitem{Machacek:1983tz}
M.~E. Machacek and M.~T. Vaughn, {\it {Two Loop Renormalization Group Equations
  in a General Quantum Field Theory. 1. Wave Function Renormalization}},  {\em
  Nucl.Phys.} {\bf B222} (1983) 83.

\bibitem{Arason:1991ic}
H.~Arason, D.~Castano, B.~Keszthelyi, S.~Mikaelian, E.~Piard, et~al., {\it
  {Renormalization group study of the standard model and its extensions. 1. The
  Standard model}},  {\em Phys.Rev.} {\bf D46} (1992) 3945--3965.

\bibitem{Sperling:2013eva}
M.~Sperling, D.~Stöckinger, and A.~Voigt, {\it {Renormalization of vacuum
  expectation values in spontaneously broken gauge theories}},  {\em JHEP} {\bf
  1307} (2013) 132, [\href{http://xxx.lanl.gov/abs/1305.1548}{{\tt
  arXiv:1305.1548}}].

\bibitem{HiggsNonBasis}
{\it {LHC Higgs Cross Section Working Group 2}},
  \href{http://xxx.lanl.gov/abs/, LHCHXSWG-INT-2015-001}{{\tt ,
  LHCHXSWG-INT-2015-001}}.

\bibitem{Falkowski:2015fla}
A.~Falkowski, {\it {Effective field theory approach to LHC Higgs data,
  arXiv:1505.00046}}, .

\bibitem{Denner:1991kt}
A.~Denner, {\it {Techniques for calculation of electroweak radiative
  corrections at the one loop level and results for W physics at LEP-200}},
  {\em Fortsch.Phys.} {\bf 41} (1993) 307--420,
  [\href{http://xxx.lanl.gov/abs/0709.1075}{{\tt arXiv:0709.1075}}].

\bibitem{Chiu:2009mg}
J.-y. Chiu, A.~Fuhrer, R.~Kelley, and A.~V. Manohar, {\it {Factorization
  Structure of Gauge Theory Amplitudes and Application to Hard Scattering
  Processes at the LHC}},  {\em Phys.Rev.} {\bf D80} (2009) 094013,
  [\href{http://xxx.lanl.gov/abs/0909.0012}{{\tt arXiv:0909.0012}}].

\bibitem{Passarino:1978jh}
G.~Passarino and M.~Veltman, {\it {One Loop Corrections for e+ e- Annihilation
  Into mu+ mu- in the Weinberg Model}},  {\em Nucl.Phys.} {\bf B160} (1979)
  151.

\bibitem{Vermaseren:2000nd}
J.~Vermaseren, {\it {New features of FORM}},
  \href{http://xxx.lanl.gov/abs/math-ph/0010025}{{\tt math-ph/0010025}}.

\bibitem{Hahn:1998yk}
T.~Hahn and M.~Perez-Victoria, {\it {Automatized one loop calculations in
  four-dimensions and D-dimensions}},  {\em Comput.Phys.Commun.} {\bf 118}
  (1999) 153--165, [\href{http://xxx.lanl.gov/abs/hep-ph/9807565}{{\tt
  hep-ph/9807565}}].

\bibitem{Mertig:1990an}
R.~Mertig, M.~Bohm, and A.~Denner, {\it {FEYN CALC: Computer algebraic
  calculation of Feynman amplitudes}},  {\em Comput.Phys.Commun.} {\bf 64}
  (1991) 345--359.

\bibitem{Hahn:2004fe}
T.~Hahn, {\it {CUBA: A Library for multidimensional numerical integration}},
  {\em Comput.Phys.Commun.} {\bf 168} (2005) 78--95,
  [\href{http://xxx.lanl.gov/abs/hep-ph/0404043}{{\tt hep-ph/0404043}}].

\bibitem{Hagiwara:1993qt}
K.~Hagiwara, R.~Szalapski, and D.~Zeppenfeld, {\it {Anomalous Higgs boson
  production and decay}},  {\em Phys.Lett.} {\bf B318} (1993) 155--162,
  [\href{http://xxx.lanl.gov/abs/hep-ph/9308347}{{\tt hep-ph/9308347}}].

\bibitem{Manohar:1983md}
A.~Manohar and H.~Georgi, {\it {Chiral Quarks and the Nonrelativistic Quark
  Model}},  {\em Nucl.Phys.} {\bf B234} (1984) 189.

\bibitem{Jenkins:2013sda}
E.~E. Jenkins, A.~V. Manohar, and M.~Trott, {\it {Naive Dimensional Analysis
  Counting of Gauge Theory Amplitudes and Anomalous Dimensions}},  {\em
  Phys.Lett.} {\bf B726} (2013) 697--702,
  [\href{http://xxx.lanl.gov/abs/1309.0819}{{\tt arXiv:1309.0819}}].

\bibitem{Buchalla:2013eza}
G.~Buchalla, O.~Catá, and C.~Krause, {\it {On the Power Counting in Effective
  Field Theories}},  {\em Phys.Lett.} {\bf B731} (2014) 80--86,
  [\href{http://xxx.lanl.gov/abs/1312.5624}{{\tt arXiv:1312.5624}}].

\bibitem{Buchalla:2014eca}
G.~Buchalla, O.~Cata, and C.~Krause, {\it {A Systematic Approach to the SILH
  Lagrangian}},  {\em Nucl.Phys.} {\bf B894} (2015) 602--620,
  [\href{http://xxx.lanl.gov/abs/1412.6356}{{\tt arXiv:1412.6356}}].

\bibitem{Cheung:2015aba}
C.~Cheung and C.-H. Shen, {\it {Non-renormalization Theorems without
  Supersymmetry,arXiv:1505.01844}}, .

\bibitem{Alonso:2014rga}
R.~Alonso, E.~E. Jenkins, and A.~V. Manohar, {\it {Holomorphy without
  Supersymmetry in the Standard Model Effective Field Theory}},  {\em
  Phys.Lett.} {\bf B739} (2014) 95--98,
  [\href{http://xxx.lanl.gov/abs/1409.0868}{{\tt arXiv:1409.0868}}].

\bibitem{Elias-Miro:2014eia}
J.~Elias-Miro, J.~Espinosa, and A.~Pomarol, {\it {One-loop non-renormalization
  results in EFTs}},  \href{http://xxx.lanl.gov/abs/1412.7151}{{\tt
  arXiv:1412.7151}}.

\bibitem{Elias-Miro:2013mua}
J.~Elias-Miro, J.~Espinosa, E.~Masso, and A.~Pomarol, {\it {Higgs windows to
  new physics through d=6 operators: constraints and one-loop anomalous
  dimensions}},  {\em JHEP} {\bf 1311} (2013) 066,
  [\href{http://xxx.lanl.gov/abs/1308.1879}{{\tt arXiv:1308.1879}}].

\bibitem{Aad:2014eha}
{\bf ATLAS} Collaboration, G.~Aad et~al., {\it {Measurement of Higgs boson
  production in the diphoton decay channel in pp collisions at center-of-mass
  energies of 7 and 8 TeV with the ATLAS detector}},  {\em Phys.Rev.} {\bf D90}
  (2014), no.~11 112015, [\href{http://xxx.lanl.gov/abs/1408.7084}{{\tt
  arXiv:1408.7084}}].

\bibitem{Khachatryan:2014ira}
{\bf CMS} Collaboration, V.~Khachatryan et~al., {\it {Observation of the
  diphoton decay of the Higgs boson and measurement of its properties}},  {\em
  Eur.Phys.J.} {\bf C74} (2014), no.~10 3076,
  [\href{http://xxx.lanl.gov/abs/1407.0558}{{\tt arXiv:1407.0558}}].

\bibitem{Aad:2015zhl}
{\bf ATLAS, CMS} Collaboration, G.~Aad et~al., {\it {Combined Measurement of
  the Higgs Boson Mass in $pp$ Collisions at $\sqrt{s}=7$ and 8 TeV with the
  ATLAS and CMS Experiments}},  \href{http://xxx.lanl.gov/abs/1503.0758}{{\tt
  arXiv:1503.0758}}.

\bibitem{Bergstrom:1985hp}
L.~Bergstrom and G.~Hulth, {\it {Induced Higgs Couplings to Neutral Bosons in
  $e^+ e^-$ Collisions}},  {\em Nucl.Phys.} {\bf B259} (1985) 137.

\bibitem{Manohar:2006gz}
A.~V. Manohar and M.~B. Wise, {\it {Modifications to the properties of the
  Higgs boson}},  {\em Phys.Lett.} {\bf B636} (2006) 107--113,
  [\href{http://xxx.lanl.gov/abs/hep-ph/0601212}{{\tt hep-ph/0601212}}].

\bibitem{Bardin:1999ak}
D.~Y. Bardin and G.~Passarino, {\it {The standard model in the making:
  Precision study of the electroweak interactions}}, .

\bibitem{Arzt:1994gp}
C.~Arzt, M.~Einhorn, and J.~Wudka, {\it {Patterns of deviation from the
  standard model}},  {\em Nucl.Phys.} {\bf B433} (1995) 41--66,
  [\href{http://xxx.lanl.gov/abs/hep-ph/9405214}{{\tt hep-ph/9405214}}].

\bibitem{Jenkins:2013fya}
E.~E. Jenkins, A.~V. Manohar, and M.~Trott, {\it {On Gauge Invariance and
  Minimal Coupling}},  {\em JHEP} {\bf 1309} (2013) 063,
  [\href{http://xxx.lanl.gov/abs/1305.0017}{{\tt arXiv:1305.0017}}].

\bibitem{Manohar:2013rga}
A.~V. Manohar, {\it {An Exactly Solvable Model for Dimension Six Higgs
  Operators and $h \to \gamma \gamma$}},  {\em Phys.Lett.} {\bf B726} (2013)
  347--351, [\href{http://xxx.lanl.gov/abs/1305.3927}{{\tt arXiv:1305.3927}}].

\bibitem{'tHooft:1971fh}
G.~'t~Hooft, {\it {Renormalization of Massless Yang-Mills Fields}},  {\em
  Nucl.Phys.} {\bf B33} (1971) 173--199.

\bibitem{Faddeev:1967fc}
L.~Faddeev and V.~Popov, {\it {Feynman Diagrams for the Yang-Mills Field}},
  {\em Phys.Lett.} {\bf B25} (1967) 29--30.

\bibitem{Gupta:2014rxa}
R.~S. Gupta, A.~Pomarol, and F.~Riva, {\it {BSM Primary Effects}},  {\em
  Phys.Rev.} {\bf D91} (2015), no.~3 035001,
  [\href{http://xxx.lanl.gov/abs/1405.0181}{{\tt arXiv:1405.0181}}].

\bibitem{Gaillard:1985uh}
M.~Gaillard, {\it {The Effective One Loop Lagrangian With Derivative
  Couplings}},  {\em Nucl.Phys.} {\bf B268} (1986) 669.

\bibitem{Drozd:2015kva}
A.~Drozd, J.~Ellis, J.~Quevillon, and T.~You, {\it {Comparing EFT and Exact
  One-Loop Analyses of Non-Degenerate Stops, arXiv:1504.02409}}, .

\end{thebibliography}\endgroup

\end{document}